\documentclass[11pt,epsf]{article}
\usepackage{amsmath,amssymb}
\usepackage{epsfig}
\usepackage{color}

\newcommand{\prt}{\partial}

\makeatletter
\usepackage[colorlinks=true,citecolor=blue]{hyperref}
\usepackage{fullpage}
\usepackage{mathpazo,times}
\def\bse{\begin{subequations}}
\def\ese{\end{subequations}}
\numberwithin{equation}{section}
\advance\parskip2pt
\advance\textheight 8mm
\renewcommand\section{\@startsection {section}{1}{\z@}%
  {-2.6ex \@plus -1ex \@minus -.2ex}%
  {1.6ex \@plus.1ex}%
  {\normalfont\bf\sffamily}}
\renewcommand\subsection{\@startsection{subsection}{2}{\z@}%
  {-2ex\@plus -0.4ex \@minus -.2ex}%
  {1.2ex \@plus .1ex}%
  {\normalfont\small\bf\sffamily}}
\renewcommand\subsubsection{\@startsection{subsubsection}{3}{\z@}%
  {-0.6ex\@plus -0.2ex \@minus -.2ex}%
  {0.4ex \@plus .1ex}%
  {\normalfont\normalsize\it}}
\renewcommand\paragraph{\@startsection{paragraph}{4}{\z@}%
  {0.2ex \@plus0.2ex \@minus0.1ex}{-0.5em}%
  {\normalfont\normalsize\bfseries}}
\allowdisplaybreaks
\makeatother

\def\be{\begin{equation}}
\def\ee{\end{equation}}

\def\Xint#1{\mathchoice
   {\XXint\displaystyle\textstyle{#1}}%
   {\XXint\textstyle\scriptstyle{#1}}%
   {\XXint\scriptstyle\scriptscriptstyle{#1}}%
   {\XXint\scriptscriptstyle\scriptscriptstyle{#1}}%
   \!\int}
\def\XXint#1#2#3{{\setbox0=\hbox{$#1{#2#3}{\int}$}
     \vcenter{\hbox{$#2#3$}}\kern-.5\wd0}}

\def\dashint{\Xint-}

\begin{document}
\bibliographystyle{plain}

\title{{\bf\sffamily Whitham modulation theory for (2+1)-dimensional equations of Kadomtsev-Petviashvili type}}
\author{Mark~J.~Ablowitz\footnote{Department of Applied Mathematics, University of Colorado, Boulder, CO 80309 },
 Gino Biondini\footnote{Department of Mathematics, State University of New York at Buffalo, Buffalo, New York 14260-2900}
 ~and 
 Igor~Rumanov\footnote{corresponding author; e-mail: igor.rumanov@colorado.edu; Department of Applied Mathematics, University of Colorado, Boulder, CO 80309} 
}
\date{\normalsize\today}

\maketitle

\kern-4\medskipamount
\begin{abstract}
Whitham modulation theory for certain two-dimensional evolution equations of Kadomtsev-Petvia\-shvili (KP) type is presented. 
Three specific examples are considered in detail: the KP equation, the two-dimensional Benjamin-Ono (2DBO) equation and a modified KP (m2KP) equation. A unified derivation is also provided. 
In the case of the m2KP equation, the corresponding Whitham modulation system exhibits features different from the other two.
~The approach presented here does not require integrability of the original evolution equation.  
Indeed, while the KP equation is known to be a completely integrable equation, the 2DBO equation and the m2KP equation are not known to be integrable. 
In each of the cases considered,
the Whitham modulation system obtained consists of five first-order quasilinear partial differential equations.
The Riemann problem (i.e.~the analogue of the Gurevich-Pitaevskii problem)
for the one-dimensional reduction of the m2KP equation is studied.
For the m2KP equation, the system of modulation equations is used to analyze the linear stability of traveling wave solutions.
\end{abstract}

\bigskip

\section{Introduction and main results}
\label{s:main}

Beginning with the seminal work of Whitham~\cite{Whitham65}, small dispersion limits and dispersive shock waves (DSWs) have been intensely studied.
Using the framework of Whitham modulation theory, Gurevich and Pitaevskii~\cite{GurPit73} analyzed the Korteweg-de Vries (KdV) equation
and found a physically important solution of the modulation equations corresponding to the evolution of step initial conditions. This solution is described by a rapidly varying cnoidal function with a modulated envelope, which today is referred to as a DSW. 
Many interesting  analytic and numerical results have been obtained over the intervening years, see e.g.~the recent reviews~\cite{ElHoefRev16, Grava17} and references therein. The vast majority of the results obtained relate to (1+1)-dimensional partial differential equations (PDEs) such as the KdV equation, the nonlinear Schr\"odinger (NLS) and the Benjamin-Ono (BO) equation.

Until recently, relatively few corresponding results had been available for (2+1)-dimensional PDEs. 
Whitham modulation theory for (2+1)-dimensional {\it integrable} PDEs and the KP equation in particular was presented in~\cite{Kr88}.
The approach was based on spectral properties of Baker-Akhiezer wavefunctions of the associated Lax pairs.
There also were early derivations of Whitham systems for the KP equation along the lines of the original Whitham approach, see~\cite{Bog91, InfRow}. But none of those works derived associated hydrodynamic systems, such as those found in \cite{GurPit73}, 
which is a crucial component in the development of the theory. 
For further details on this issue, we refer the reader to the discussion at the end of section 2 of~\cite{ABW-KP}. 

Recently, however, several studies have been devoted to Whitham theory for (2+1)-dimensional systems.
In particular, \cite{ADM} demonstrated the derivation and physical relevance of the Whitham systems for certain reductions of the KP and two-dimensional Benjamin-Ono (2DBO) equations. These reductions lead to the cylindrical KdV and cylindrical BO equations, hence they are quite different from the standard  (1+1)-dimensional KdV and BO reductions of the KP and 2DBO equations, respectively. 
From the point of view of Whitham theory, the reductions of these (2+1)-dimensional PDEs  can be considered on the same footing as the (1+1)-dimensional ones. 
Subsequently Whitham theory has been considered for the (2+1) dimensional KP~\cite{ABW-KP} and 2DBO~\cite{ABW-BO} without 
using any one-dimensional reductions.  We also mention that the underlying structure of solutions that will develop dispersive shocks in the small dispersion limit of the generalized
KP equations has been studied in~\cite{DuGrKl16}. 
We refer the reader to these papers for additional background and references.

In this work we present an approach that
applies equally well to integrable and non-integrable PDEs, since at the heart of it lies Whitham's WKB type expansion and separation of fast and slow scales. 
We consider (2+1)-dimensional PDEs of the form 
\be
\prt_x[\prt_tu + F(u, \prt_xu, \dots, \prt_x^nu; \epsilon)] + \alpha \prt_{yy}u = 0.   
\label{eq:1.1}
\ee
We refer to these PDEs as KP-type equations because of the common $y$-dependence and the functional $F(u, \prt_xu, \dots, \prt_x^nu; \epsilon)$ employed, which consists of one nonlinear convective term and one dispersive term.
Here $\epsilon$ is a dispersion parameter, which is assumed to be small. We show here that the Whitham modulation systems for PDEs of this form can be derived in a unified way, which also clarifies and simplifies the derivations given in~\cite{ABW-KP, ABW-BO}. The examples we treat in this paper are the following:
\bse
\begin{enumerate}
\item 
The KP equation corresponds to 
\begin{equation}
F(u, \prt_xu, \dots, \prt_x^nu; \epsilon) = 6u\prt_xu + \epsilon^2\prt_{xxx}u\,.    
\end{equation}
The above equation is completely integrable and has been widely studied. It is an important model in the study of nonlinear dispersive waves, and arises in many physical contexts including surface water waves, plasmas, ferromagnetics and cosmology. 
The case $\alpha=1$ is known as the KP II equation, and describes water waves with small surface tension, while the case $\alpha=-1$ is called the KP I equation, and describes water waves with strong surface tension~\cite{AbSeg81,InfRow}.
\item 
The 2DBO equation has  
\begin{equation}
F(u, \prt_xu, \dots, \prt_x^nu; \epsilon) = u\prt_xu + \epsilon\mathcal H[\prt_{xx}u]\,,    
\end{equation}
where $\mathcal H[f]$ denotes the Hilbert transform of $f$ (see Appendix~A for the definition). 
This equation describes two-dimensional internal waves in stratified fluids \cite{AbSeg80}.
\item 
The modified Kadomtsev-Petviashvili (m2KP) equation is associated with
\begin{equation}
F(u, \prt_xu, \dots, \prt_x^nu; \epsilon) = -6u^2\prt_xu + \epsilon^2\prt_{xxx}u\,.    
\end{equation}
This equation arises in the description of sound waves in anti-ferromagnets \cite{TuFa85}.
Note that this m2KP equation should not be confused with an integrable modified KP equation~\cite{Kon82, JM83} which is also related to the (1+1)-dimensional modified KdV (mKdV) equation. The mKdV equation 
is the case $\alpha=0$ of the m2KP equation considered here; this m2KP equation is not known to be integrable. It is also the $n=2$ member (in the defocusing case) of the generalized KP equations with a nonlinear term in $F$ of the form $\pm6u^n\prt_xu$ considered in~\cite{TuFa85, WAS94}, see also~\cite{DuGrKl16} for a recent review. Each of these generalized KP equations also belongs to the class of KP type equations (\ref{eq:1.1}) and, in principle, could be treated along the same lines as the three examples above.
\end{enumerate}
\ese

Importantly, all the evolution equations of the type~\eqref{eq:1.1} can also be written as systems of the form
\vspace*{-1ex}
\bse
\label{e:evolutionsystem}
\begin{gather}
\prt_tu + F(u, \prt_xu, \dots, \prt_x^nu; \epsilon) + \alpha \prt_yv = 0,   
\label{eq:u}
\\
\prt_xv = \prt_yu.   
\label{eq:v}
\end{gather}
\ese
We look for solutions of the system~\eqref{e:evolutionsystem} depending on fast and slow scales, namely  
\be
u = u(\theta, x, t; \epsilon),   \qquad  v = v(\theta, x, t; \epsilon), 
\label{eq:2.1}
\ee
where we impose the following conditions for the fast phase $\theta$:
\be
\prt_x\theta = \frac{k}{\epsilon},  \qquad \prt_y\theta = \frac{l}{\epsilon},  \qquad \prt_t\theta = -\frac{\omega}{\epsilon},  \label{eq:2.2}
\ee
with $0<\epsilon \ll 1$, and where $k$, $l$ and $\omega$ are slowly varying quantities. 
Equations~\eqref{eq:2.2} imply immediately
\be
\prt_tk+\prt_x\omega = 0,  \qquad  \prt_tl + \prt_y\omega=0,   \qquad   \prt_yk=\prt_xl.   
\label{eq:klo}
\ee
The first two equations are referred to as conservation of waves, 
and will provide the first and second modulation equations, 
while the third equation is a constraint that will be used in the derivation.  
It is convenient to
introduce the notations
\be
q=l/k, \qquad V = \omega/k - \alpha q^2,   \label{e:qVdef}
\ee
as well as the convective derivative
\be
D_y = \prt_y-q\prt_x\,.
\ee
Then Eqs.~(\ref{eq:klo}) can be rewritten respectively as
\bse
\begin{gather}
\prt_tk + \prt_x(k(V+\alpha q^2)) = 0,   \label{eq:k}
\\
\prt_t(kq) + \prt_y(k(V+\alpha q^2)) = 0,  \label{eq:l}
\\
\frac{D_yk}{k} = \prt_xq.   \label{eq:kl}
\end{gather}
\ese
Next we substitute $\prt_tk$ expressed from Eq.~(\ref{eq:k}) into Eq.~(\ref{eq:l}), and, using also Eq.~(\ref{eq:kl}), obtain the Whitham modulation equation which determines the evolution of $q$, namely
\be
\prt_tq + (V+\alpha q^2)\prt_xq + D_y(V+\alpha q^2) = 0.  
\label{eq:q}
\ee
Eqs.~(\ref{eq:k}), (\ref{eq:kl}) and (\ref{eq:q}) are the part of the Whitham modulation equations that is common to all examples considered here. 
These ``kinematic" equations remain intact at all higher orders in $\epsilon$, since they are just a consequence of the definition of the fast phase $\theta$ via Eq.~\eqref{eq:2.2}.

Next we expand $u=u(\theta,x,y,t)$ and $v(\theta,x,y,t)$ in series in $\epsilon$, namely,
\be
u = u_0 + \epsilon u_1 + O(\epsilon^2),  
\qquad  
v = v_0 + \epsilon v_1 + O(\epsilon^2)\,
\label{e:expansion}
\ee
and introduce the multiple scales 
\be
\partial_x \to (k/\epsilon)\partial_\theta + \partial_x,  \qquad
\partial_y \to (l/\epsilon)\partial_\theta + \partial_y,    \qquad
\partial_t  \to -(\omega/\epsilon)\partial_\theta + \partial_t.
\label{mscales}
\ee
We substitute Eqs.~\eqref{e:expansion}-\eqref{mscales} into the system~\eqref{e:evolutionsystem}. 
Requiring that secular terms be absent in the expansion then yields the remaining three modulation equations. 
The general formalism for the derivation of these equations is discussed in section~\ref{s:derivation}.
On the other hand, one of the remaining three modulation equations also has a universal form, as we show next.

We denote $\prt_{\theta}f = f'$ for brevity. At leading order [i.e., $O(1/\epsilon)$] of Eq.~\eqref{eq:v} we have
\vspace*{-1ex}
\be
kv_0' - qk u_0' = 0\,.
\label{eq:leadingordersystem2}
\ee
Equation~\eqref{eq:leadingordersystem2} is readily solved to obtain
\be
v_0 = qu_0 + p,  
\label{eq:v0}
\ee
where the integration ``constant'' $p=p(x,y,t)$ is a slow variable to be determined at the next order in the expansion. At the next order [i.e., $O(1)$] of Eq.~\eqref{eq:v} we have 
\vspace*{-1ex}
\be
kv_1' - qk u_1' = \prt_yu_0 - \prt_x(qu_0 + p)\,.
\label{eq:O1system2}
\ee
The Whitham equations can be derived as secularity conditions ensuring that the corrections $u_1$ and $v_1$ to the leading order solution are periodic rather than growing in $\theta$. Let $\overline f$ denote the 
average of a periodic function $f(\theta)$ over its period, which, without loss of generality, we can fix to be $1$; i.e.
$$
\overline f = \int_0^1f(\theta)d\theta\,.
$$
Imposing periodicity in $\theta$ of the functions involved and integrating eq.~(\ref{eq:O1system2}) over the period leads to the secularity condition
\be
\partial_xp = D_y\overline{u_0} - \overline{u_0}\partial_xq.   
\label{eq:S3}
\ee
This Whitham equation is also common to all PDEs of the KP type. The form of the remaining two Whitham equations depends on the specifics of the function $F$ in Eq.~(\ref{eq:u}), as discussed in section~\ref{s:derivation}.

In sections~\ref{s:KP}--\ref{s:m2KP} we show in detail how, in the three specific examples considered here, 
this method yields the complete Whitham modulation systems for the KP, 2DBO and m2KP equations. 
We also show how, in all three cases, one can express each of these modulation equations in terms of the three known Riemann invariants 
$r_1,r_2,r_3$ 
of the corresponding (1+1)-dimensional Whitham system, together with the additional variables $q$ and $p$ introduced above. 
In this way one can ``diagonalize'' the evolution equations for the dependent variables other than $q$ and $p$ with respect to these Riemann variables in exactly the same way as this would have been done for its (1+1)-dimensional counterpart.

The current method simplifies the earlier treatment of the KP and 2DBO equations in~\cite{ABW-KP,ABW-BO}. In particular, it turns out that the singularity considerations in~\cite{ABW-KP, ABW-BO} are not necessary. 
The Whitham Eq.~(\ref{eq:q}) remains intact, with only the velocity $V$ in it expressed in terms of the Riemann type variables. 
Equation~\eqref{eq:S3} also remains intact up to expressing $\overline{u_0}$ in Riemann variables (and optionally using Eq.~(\ref{eq:kl}) in it). 
The only equations subject to transformation are Eq.~\eqref{eq:k} and the two secularity equations specific to each case (see sections~\ref{s:KP}--\ref{s:m2KP}), which, however, are also derived in a unified way for KP-type systems, see  Eqs.~\eqref{e:Whitham4} and~\eqref{e:Whitham5} in section~\ref{s:derivation}.
In other words, e.g.~for the case of the KP equation, we transform only the counterpart of the corresponding KdV-Whitham equations to make their KdV parts 
diagonal in terms of the Riemann variables of the KdV-Whitham system. 
For the KP equation, a similar approach was recently presented in~\cite{GrKlPi17}, 
where four modulation equations were derived, and no analogue of the dependent variable $p$ was introduced.
The system in~\cite{GrKlPi17} can be obtained from the system~\eqref{e:KP-Whitham} or that of~\cite{ABW-KP} if $p(x,y,t)=0$ and Eq.~(\ref{eq:S3}) is omitted. 
We emphasize, however, that, \textit{as shown in~\cite{ABW-KP}, in order to correctly describe the dynamics of the original (2+1)-dimensional evolution equation,
non-trivial values of $p(x,y,t)$ must be considered, and equation Eq.~(\ref{eq:S3}) is therefore required}. 
The same considerations are valid for the other two evolution equations mentioned above (namely, the 2DBO and m2KP equations), and indeed apply to any equation of the type~\eqref{eq:1.1}.

Following this approach, in sections~\ref{s:KP}--\ref{s:m2KP} we obtain
`hydrodynamic' Whitham systems for the three cases at hand in the final form presented below, 
consisting in each case of five quasi-linear first-order PDEs for the dependent variables $r_1,r_2,r_3,q,p$ 
as functions of the slow coordinates $x,y$ and $t$. In each case, the Riemann variables $r_1,r_2,r_3$ are introduced in 
an identical manner as in the corresponding one-dimensional case,
while the variable $p$ is the part of $v_0$ that is independent of $\theta$ and depends only on the slow coordinates. Specifically, we have:
%
\begin{enumerate}
\item The KP-Whitham system:
\bse
\label{e:KP-Whitham}
\begin{gather}
\prt_tr_j + (v_j + \alpha q^2)\prt_xr_j + 2\alpha qD_yr_j + \alpha\left(2r_j - \frac{v_j}{6}\right)D_yq + \alpha D_yp = 0,   \qquad j=1,2,3,
\label{eq:rj}
\\
\prt_tq + (V+\alpha q^2)\prt_xq + D_y(V+\alpha q^2) = 0,  \label{eq:qKP}
\\
\prt_xp - \bigg(1 - \frac{E}{K}\bigg)D_yr_1 - \frac{E}{K}D_yr_3 + (r_1-r_2+r_3)\prt_xq = 0.  \label{eq:pKP}
\end{gather}
\ese
where $V = 2(r_1+r_2+r_3)$ and
$v_j$ are the well-known characteristic velocities for the KdV-Whitham system~\cite{Whitham65, Whitham74}, namely
\bse
\be
v_j = V + \frac{2}{\prt_jk/k}, 
\label{eq:vj}
\ee
where $\prt_jk/k \equiv {\prt \ln k}/{\prt r_j}$ are logarithmic derivatives of $k$ with respect to the Riemann invariants $r_1,r_2,r_3$, given by 
\be
\frac{\prt_1k}{k} = -\frac{(1 - E/K)}{2(r_2-r_1)},   \quad  \frac{\prt_2k}{k} = \frac{1}{2}\left( \frac{1-E/K}{r_2-r_1} - \frac{E/K}{r_3-r_2} \right),  \quad  \frac{\prt_3k}{k} = \frac{E/K}{2(r_3-r_2)},   \label{eq:djk}
\ee
\ese
and $K=K(m)$, $E=E(m)$ are the first and second complete elliptic integrals, respectively, with $m=(r_2-r_1)/(r_3-r_1)$ (see Appendix~A for details).
The system~\eqref{e:KP-Whitham} is equivalent to the KP-Whitham system derived in \cite{ABW-KP}, as discussed at the end of section~2.

\item 
The 2DBO-Whitham system:
\bse
\label{e:BO-Whitham}
\begin{gather}
\prt_tr_j + 2r_j\prt_xr_j + \alpha\left(q^2\prt_xr_j + 2qD_yr_j + c_jD_yq + {D_yp}\right) = 0,   \qquad j=1,2,3,
\label{eq:r1}   
\\
\prt_tq + (V+\alpha q^2)\prt_xq + D_y(V+\alpha q^2) = 0,  \label{eq:qBO}
\\
\prt_xp = D_yr_1 - r_1\prt_xq\,,
\label{eq:pBO}
\end{gather}
\ese
with $V = r_3 + r_2$, $c_1=r_3-r_2+r_1$, $c_2=c_3 = r_3+r_2-r_1$ and where we redefined $p\mapsto 2p$ for simplicity.
The system~\eqref{e:BO-Whitham} is equivalent to the the 2DBO system in \cite{ABW-BO}, as discussed at the end of section 3. 

\item 
The m2KP-Whitham system:
\bse
\label{e:m2KP-Whitham}
\begin{gather}
\prt_t r_j + (v_j + \alpha q^2)\prt_x r_j + 2\alpha qD_y r_j 
\kern24em{ }
\nonumber\\
\kern4em{ }
+ \alpha\frac{r_j\left[ ( r_jQ_2 -  r_i r_lQ)D_yq + ( r_jQ- r_i r_l)D_yp \right]}{( r_j^2- r_i^2)( r_j^2- r_l^2)\prt_jk/k} = 0,   \qquad j=1,2,3,
\label{eq:Rj}
\\
\prt_tq + (V+\alpha q^2)\prt_xq + D_y(V+\alpha q^2) = 0,  \label{eq:mq}
\\
\prt_xp + \sum_{j=1}^3\frac{ r_i r_l}{ r_j}\frac{\prt_jk}{k}\; D_y r_j = 0,  \label{eq:mp}
\end{gather}
\ese
with $j\neq i\neq l\neq j$, where $V = -2( r_1^2 + r_2^2 + r_3^2)$ and $v_j$ are the characteristic velocities for the mKdV-Whitham system~\cite{DN-mKdV}, namely
\bse
\label{e:m2KPcoeffs}
\be
v_j = V - \frac{4 r_j}{\prt_jk/k}\,,
\label{eq:mvj}
\ee
where $\prt_jk/k \equiv {\prt \ln k}/{\prt r_j}$ are logarithmic derivatives of $k$ with respect to the Riemann invariants $r_1, r_2, r_3$, given by 
\be
\frac{\prt_1k}{k} = -\frac{ r_1\; E/K}{ r_2^2- r_1^2},   \quad  \frac{\prt_2k}{k} = r_2\left( \frac{E/K}{ r_2^2- r_1^2} - \frac{1-E/K}{ r_3^2- r_2^2} \right),  \quad  \frac{\prt_3k}{k} = \frac{ r_3(1-E/K)}{ r_3^2- r_2^2},   
\label{eq:kjR}
\ee
with $K=K(m)$ and $E=E(m)$ as before,
\begin{gather}
Q = r_2 + r_3 - r_1 - 2( r_2- r_1)\frac{\Pi}{K},   \qquad  Q_2 = r_2^2 + r_3^2 - r_1^2 - 2(r_3^2- r_1^2)\frac{E}{K},    \label{eq:QR}
\\
k^2 = \frac{ r_3^2- r_1^2}{4K},  \qquad \gamma = \frac{ r_3- r_2}{ r_3- r_1},   \quad m = \frac{ r_3^2- r_2^2}{ r_3^2- r_1^2},   \label{eq:mgR}
\end{gather}
\ese
and where $\Pi = \Pi(\gamma,m)$ is the third complete elliptic integral (see Appendix~A for details). 
The above m2KP-Whitham system~\eqref{e:m2KP-Whitham} is new. 
\end{enumerate}
Note that while the above systems of PDEs are closed and can be studied in their own right, if one wants to use them to study the behavior of solutions of the original PDEs in the small dispersion limit, one has to add to each of them the important constraint Eq.~\eqref{eq:kl}.

The outline of the remainder of this work is the following. 
In sections~\ref{s:KP} and~\ref{s:BO} we describe the derivation of the Whitham systems for the KP and 2DBO equations, respectively, which, as mentioned above, simplifies the derivation in~\cite{ABW-KP, ABW-BO}. 
Section~\ref{s:m2KP} describes the derivation of the Whitham system for the m2KP equation, which has novel features.
Section~\ref{s:m2KPstability} is devoted to applications of the m2KP-Whitham equations.
Namely, we study the linear stability of periodic solutions of the m2KP equation using the m2KP-Whitham system~\eqref{e:m2KP-Whitham}.
This is similar to what was done for the KP equation in~\cite{ABW-KP} and the 2DBO equation in~\cite{ABW-BO}. 
Interestingly, however, in the case of the m2KP equation the (in)stability picture turns out to be richer than in the previous two cases. 
Specifically, the stability properties of the cnoidal solutions of the m2KP equation depend on two parameters rather than one. Moreover,
both linear spectral stability and instability can occur for each sign of constant $\alpha$.
In section~\ref{s:derivation} we present the general derivation of the Whitham modulation equations for equations in the form~\eqref{eq:1.1}, and 
section~\ref{s:conclusions} offers some concluding remarks. Appendices~A--D contain related material.
In particular, Appendices~B and~C discuss the diagonalization of the Whitham systems for the KP and m2KP equations respectively,
and in Appendix~D we discuss the analogue of the Gurevich-Pitaevskii problem for the defocusing mKdV.

\section{Derivation of the KP-Whitham system}
\label{s:KP}

\subsection{Modulation equations}

The Kadomtsev-Petviashvili (KP) equation can be written as the evolution system~\eqref{e:evolutionsystem}, where in particular Eq.~\eqref{eq:u} takes the form
\bse
\label{e:KPevolutionsystem}
\be
\prt_tu + 6u\prt_xu + \epsilon^2\prt_{xxx}u + \alpha \prt_yv = 0,   \label{eq:uKP0}
\ee
while the common Eq.~(\ref{eq:v}), which we repeat for convenience, is
\be
\prt_xv = \prt_yu.   \label{eq:vKP0}
\ee
\ese
After introducing fast and slow scales, the system~\eqref{e:KPevolutionsystem} becomes
\bse
\begin{gather}
\left(-\frac{\omega}{\epsilon}\prt_{\theta} + \prt_t\right)u + 6u\left(\frac{k}{\epsilon}\prt_{\theta} + \prt_x\right)u + \epsilon^2\left(\frac{k}{\epsilon}\prt_{\theta} + \prt_x\right)^3u + \alpha \left(\frac{kq}{\epsilon}\prt_{\theta} +\prt_y\right)v = 0,   \label{eq:uKP}
\\
\left(\frac{k}{\epsilon}\prt_{\theta} + \prt_x\right)v = \left(\frac{kq}{\epsilon}\prt_{\theta} +\prt_y\right)u.   \label{eq:vKP}
\end{gather}
\ese
The cubed operator in the third (dispersive) term of Eq.~(\ref{eq:uKP}) expands in powers of $\epsilon$ as
$$
\epsilon^2\left(\frac{k}{\epsilon}\prt_{\theta} + \prt_x\right)^3 = \frac{k^3}{\epsilon}\prt_{\theta}^3 + 3k^2\prt_{\theta}^2\prt_x + 3k\prt_xk\prt_{\theta}^2 + \epsilon(3k\prt_{\theta}\prt_x^2 + 3\prt_xk\prt_{x\theta} + \prt_{xx}k\prt_{\theta} ) + \epsilon^2\prt_x^3.
$$
Eqs.~(\ref{eq:k}), (\ref{eq:q}) and (\ref{eq:kl}) are the first three Whitham equations for the KP equation. 
After substituting Eq.~(\ref{eq:v0}) into Eq.~(\ref{eq:uKP}) the last equation yields, at leading order,
\be
-kVu_0' + 6ku_0u_0' + k^3u_0''' = 0.   \label{eq:u0KP}
\ee
Integrating Eq.~\eqref{eq:u0KP} once, one gets
\be
-Vu_0 + 3u_0^2 + k^2u_0'' = C_1.  \label{eq:I1}
\ee
Multiplying Eq.~(\ref{eq:I1}) by $2u_0'$ and integrating again, one obtains the well-known equation for the elliptic (``genus one'') solution of KdV,
\be
k^2(u_0')^2 = -2u_0^3 + Vu_0^2 + 2C_1u_0 + 2C_2 =-2(u_0-\lambda_1)(u_0-\lambda_2)(u_0-\lambda_3).   \label{eq:g1} 
\ee
Its general solution can be written as
\be
u_0 = a + b\;\text{cn}^2\left(2K(m)(\theta - \theta_*); m\right),   \label{eq:u0}
\ee
where $\theta_*$ is an integration constant, 
\be
a = \lambda_2,  \qquad b = \lambda_3-\lambda_2,  \qquad  m = \frac{\lambda_3-\lambda_2}{\lambda_3-\lambda_1},   \label{eq:abm}
\ee
$\lambda_1\le\lambda_2\le\lambda_3$ are the roots of the cubic in Eq.~(\ref{eq:g1}) and we have the following relations among the parameters (slow variables):
\be
\frac{V}{2} = e_1 \equiv \lambda_1 + \lambda_2 + \lambda_3,  \qquad  -C_1=e_2 \equiv \lambda_1\lambda_2+\lambda_2\lambda_3+\lambda_3\lambda_1,  \qquad C_2=e_3\equiv \lambda_1\lambda_2\lambda_3.  \label{eq:e123}
\ee
The normalization of the elliptic function with fixed period one implies that
\be
k^2 = \frac{b}{8mK^2(m)} = \frac{\lambda_3-\lambda_1}{8K^2(m)},   \label{eq:k2}
\ee
where $K(m)$ is the first complete elliptic integral. 
\par At first order in $\epsilon$ we find, after substituting Eq.~(\ref{eq:v0}) and the expression for $v_1'$ from Eq.~(\ref{eq:O1system2}) into Eq.~(\ref{eq:uKP}),
\begin{multline}
k(k^2u_1'' + 6u_0u_1 - V u_1)' +    
\\
+ \prt_tu_0 + 6u_0\prt_xu_0 + 3k^2\prt_xu_0'' + 3k\prt_xku_0'' + \alpha\left( q^2\prt_xu_0 + 2qD_yu_0 + u_0D_yq + D_yp \right) = 0.   \label{eq:u1}
\end{multline}
The other Whitham equations can be derived as secularity conditions ensuring that the solutions $u_1$ and $v_1$ of Eqs.~(\ref{eq:u1}), (\ref{eq:O1system2}) are periodic; i.e.~not growing in $\theta$. Imposing periodicity and integrating Eqs.~(\ref{eq:u1}) and (\ref{eq:O1system2}) over the period in $\theta$ leads to two secularity conditions, respectively, 
\be
\prt_t\overline{u_0} + 3\prt_x(\overline{u_0^2}) + \alpha\left( q^2\prt_x\overline{u_0} + 2qD_y\overline{u_0} + \overline{u_0}D_yq + D_yp \right) = 0,   \label{eq:S1}
\ee
and Eq.~(\ref{eq:S3}). The third (and the last needed) secularity condition is readily obtained when one notices that the terms depending on $u_1$ in Eq.~(\ref{eq:u1}) are the same as for KdV (and $v_1$ is absent in it). So we multiply Eq.~(\ref{eq:u1}) by $u_0$ and integrate over the period to find
\be
\prt_t\overline{u_0^2} + 4\prt_x\overline{u_0^3} - 3\prt_x(k^2\overline{(u_0')^2}) + \alpha\left( q^2\prt_x\overline{u_0^2} + 2D_y(q\overline{u_0^2}) + 2\overline{u_0}D_yp \right) = 0.  \label{eq:S2}
\ee
Define
\be
Q_n \equiv \int_0^1(u_0(\theta))^nd\theta = \overline{u_0^n}\,.
\label{eq:Qn}
\ee
We have
\be
Q \equiv Q_1 = a + b\left(\frac{E(m)}{mK(m)} - \frac{1-m}{m}\right) = \lambda_1 + (\lambda_3-\lambda_1)\frac{E}{K},   
\label{eq:Q1}
\ee
where $K=K(m)$ and $E=E(m)$ are the first and second complete elliptic integrals, respectively. 
Integrating Eq.~(\ref{eq:I1}) we find
\be
Q_2 \equiv \overline{u_0^2} = \frac{VQ+C_1}{3}.   \label{eq:Q2}
\ee
Integration of Eq.~(\ref{eq:g1}) gives the relation
\be
G_1 \equiv k^2\overline{(u_0')^2} = -2Q_3 + VQ_2 + 2C_1Q + 2C_2,   \label{eq:Ig1}
\ee
while multiplying Eq.~(\ref{eq:I1}) by $u_0$ and integrating, one gets
\be
-G_1 = -3Q_3 + VQ_2 + C_1Q.   \label{eq:Ig2}
\ee
Combining Eqs.~(\ref{eq:Ig1}) and (\ref{eq:Ig2}) yields 
\bse
\begin{gather}
Q_3 = \frac{1}{5}\left[ \left(\frac{2V^2}{3}+3C_1\right)Q + \frac{2VC_1}{3}+2C_2 \right],   \label{eq:Q3}
\\
G_1 = \frac{1}{5}\left[ \left(\frac{V^2}{3}+4C_1\right)Q + \frac{VC_1}{3}+6C_2 \right].   \label{eq:G1}
\end{gather}
\ese
Using Eqs.~(\ref{eq:Q2}), (\ref{eq:Q3}) and (\ref{eq:G1}), we get the secularity equations in the form ($Q \equiv \overline{u_0}$)
\bse
\begin{gather}
\prt_tQ + \prt_x(VQ+C_1) + \alpha(D_y+q\prt_x)(qQ+p) = 0,   
\label{eq:S1KP}
\\
\prt_t(VQ + C_1) + \prt_x\left(V^2Q+VC_1-6C_2\right) + \kern12em
\nonumber\\{ }\kern8em
  + \alpha\left[ q^2\prt_x(VQ+C_1) + 2D_y(q(VQ+C_1)) + 6QD_yp \right] = 0,  
\label{eq:S2KP}
\\
\prt_xp = D_yQ - Q\prt_xq.   
\label{eq:S3KP}
\end{gather}
\ese
The secularity equations (\ref{eq:S1KP}), (\ref{eq:S2KP}) and~(\ref{eq:S3KP}), plus the kinematic equations (\ref{eq:k}) and~(\ref{eq:q}) 
comprise the KP-Whitham equations in physical coordinates.

\subsection{Whitham equations in KdV Riemann variables}

The roots of the cubic in Eq.~(\ref{eq:g1}), $\lambda_i$, $i=1,2,3$, are simply related to the so-called KdV Riemann variables~\cite{Whitham65, Whitham74}
\be
\lambda_1 = r_1 + r_2 - r_3,    \qquad  \lambda_2 = r_1 - r_2 + r_3,   \qquad  \lambda_3 = -r_1 + r_2 + r_3.   \label{eq:la-r}
\ee
We are going to express the Eqs.~(\ref{eq:k}), (\ref{eq:q}), (\ref{eq:kl}), (\ref{eq:S1KP}), (\ref{eq:S2KP}) and (\ref{eq:S3KP}) in terms of Riemann variables for KdV, $r_i$, $i=1,2,3$. First, we use Eq.~(\ref{eq:e123}) and introduce the ``total" time derivative 
$$
D = \prt_t + V\prt_x,
$$ 
to rewrite Eqs.~(\ref{eq:k}), (\ref{eq:S1KP}) and (\ref{eq:S2KP}), respectively, as
\bse
\begin{gather}
\frac{1}{2}\frac{Dk}{k} + \prt_xe_1 + \frac{\alpha}{2}Y_0 = 0,   \label{eq:kKP}
\\
DQ + 2Q\prt_xe_1 - 2\prt_xe_2 + \alpha Y_1 = 0,  \label{eq:Q}
\\
DP + 2P\prt_xe_1 - 6\prt_xe_3 + \alpha Y_2 = 0,   \label{eq:P}
\end{gather}
\ese
where we denoted for further use
\begin{multline}
Y_0 = \frac{\prt_x(kq^2)}{k},  \qquad Y_1 = (D_y+q\prt_x)(qQ+p) = q^2\prt_xQ + 2qD_yQ + QD_yq + D_yp, 
\\
P \equiv VQ+C_1 = 2e_1Q-e_2,   \qquad  Y_2 = q^2\prt_xP + 2D_y(qP) + 6QD_yp.   \label{eq:Ys}
\end{multline}
Below, we transform Eqs.~(\ref{eq:kKP}), (\ref{eq:Q}) and~(\ref{eq:P}) into ``diagonal'' form in terms of the variables $r_1,r_2,r_3$.
As for Eqs.~(\ref{eq:q}), (\ref{eq:S3KP}), (\ref{eq:kl}), 
Eq.~(\ref{eq:la-r}) is used to express the functions $k$, $V$ and $Q\equiv\overline u_0$ inside them in terms of the Riemann-type $r$-variables. This leads to the final form of the KP-Whitham system.
\par Next we diagonalize the KdV parts of Eqs.~(\ref{eq:kKP}), (\ref{eq:Q}), (\ref{eq:P}) in terms of the above Riemann $r$-variables using also their power sums
\be
p_n = r_1^n + r_2^n + r_3^n.   \label{eq:pn}
\ee
The details are given in Appendix B. After this procedure, the above three Whitham equations become
\be
\prt_tr_j + v_j\prt_xr_j  + \alpha g_j = 0,   \qquad j=1,2,3,  \label{eq:rjg}
\ee
where $v_j$ are the KdV velocities (see section 1 or~Eq.~(\ref{eq:rW0}) in Appendix A) and 
\be
g_j = \frac{2}{\prt_jk/k}\frac{(r_lr_mW_1 - (r_l+r_m)W_2 + W_3)}{(r_j-r_l)(r_j-r_m)},   \qquad  j\neq l \neq m \neq j,   \label{eq:gj}
\ee
with
\bse
\begin{gather}
W_1 =  \frac{Y_0}{2} = \frac{q^2}{2}\sum_j\frac{\prt_jk}{k}\prt_xr_j + q\sum_j\frac{\prt_jk}{k}D_yr_j,  \label{eq:W1}
\\
W_2 = \frac{Y_1 + (p_1-Q)Y_0}{4} = \frac{q^2}{2}\sum_jr_j\frac{\prt_jk}{k}\prt_xr_j + q\sum_jr_j\frac{\prt_jk}{k}D_yr_j + \frac{QD_yq+D_yp}{4},   \label{eq:W2}
\\
W_3 =  \frac{1}{8}\left[ \frac{Y_2}{6} + p_1Y_1 + \frac{2}{3}(p_2+p_1^2-2p_1Q)Y_0 \right] = 
\kern16em\nonumber
\\
= \frac{q}{6}\sum_j\left[ r_j\left(4p_1\frac{\prt_jk}{k} + 1\right) + \left((p_2-p_1^2)\frac{\prt_jk}{k} + \frac{Q-p_1}{2}\right) \right]\left(\frac{q}{2}\prt_xr_j + D_yr_j\right) +
\nonumber
\\
\kern8em
+ \frac{(5p_1Q + 2p_2 - p_1^2)D_yq}{24} + \frac{(Q+p_1)D_yp}{8}.   \label{eq:W3}
\end{gather}
\ese
In order to express the quantities $W_j$ in terms of $r$-variables, we use Eqs.~(\ref{eq:QP}), (\ref{eq:dQ}) of Appendix B and Eq.~(\ref{eq:kl-r}). To simplify the quantities $g_j$ in Eq.~(\ref{eq:gj}), we use Eqs.~(\ref{eq:W1}), (\ref{eq:W2}), (\ref{eq:W3}) and (\ref{eq:Qr}) of Appendix B. Thus, after some algebra, we bring Whitham Eqs.~(\ref{eq:rjg}) to the final explicit form given by Eq.~(\ref{eq:rj}) in section~\ref{s:main}. 
The universal Eq.~(\ref{eq:q}) becomes Eq.~(\ref{eq:qKP}) upon substitution $V=2\sum_jr_j$.
\par We also use Eqs.~(\ref{eq:QP}), (\ref{eq:dQ}) of Appendix B and Eq.~(\ref{eq:kl}) in the form
\be
\sum_j\frac{\prt_jk}{k}D_yr_j = \prt_xq,   \label{eq:kl-r}
\ee
to express Eq.~(\ref{eq:S3}) as
\be
\prt_xp = \sum_j(2r_j-p_1)\frac{\prt_jk}{k}D_yr_j.   \label{eq:pKP0}
\ee
Eqs.~(\ref{eq:rj}), $j=1,2,3$, together with Eqs.~(\ref{eq:pKP0}), (\ref{eq:qKP}) obtained above, comprise the Whitham system for KP. One can verify that it is equivalent to the system in~\cite{ABW-KP}, but it is somewhat simpler. Eq.~(\ref{eq:q}) or Eq.~(\ref{eq:qKP}) here is simpler than Eq.~(1.4b) of~\cite{ABW-KP}, although they are in fact equivalent. The right-hand side of Eq.~(\ref{eq:pKP0}) is transformed, using Eq.~(\ref{eq:kl-r}) and Eq.~(\ref{eq:djk}), into Eq.~(1.4c) of~\cite{ABW-KP}, which is Eq.~(\ref{eq:pKP}) of section~\ref{s:main}. 
The Eqs.~(1.4a) of~\cite{ABW-KP} and our Eqs.~(\ref{eq:rj}) are clearly identical except for the coefficients before $D_yq$. The last, however, must match since they are the quantities appearing in the cylindrical KdV reduction. Our expression $2r_j - v_j/6$ is exactly the right form of these quantities (it would have been $2r_j - v_j$ had we taken the nonlinear term in KP as $u\prt_xu$ rather than $6u\prt_xu$, and this is verified to correspond to the terms found by~\cite{ADM}). Rearranging the terms in the corresponding Eqs.~(2.33) of~\cite{ABW-KP} one confirms that the considered coefficients indeed match. But, as mentioned in the introduction, the present approach does not introduce (removable) singularities.

\section{Derivation of the 2DBO-Whitham system}
\label{s:BO}

\subsection{Modulation equations}

The two-dimensional Benjamin-Ono (2DBO) equation can be written as the system comprised by
\be
\prt_tu + u\prt_xu + \epsilon\mathcal H[\prt_{xx}u] + \alpha \prt_yv = 0,   \label{eq:uBO0}
\ee
where $\mathcal H[f]$ is the Hilbert transform of $f$, and Eq.~(\ref{eq:v}) as before. 
Then, after introducing fast and slow scales as described in the prior section, it takes the form
\be
\left(-\frac{\omega}{\epsilon}u' + \prt_tu\right) + u\left(\frac{k}{\epsilon}u' + \prt_xu\right) + \epsilon\mathcal H\left[\frac{k^2}{\epsilon^2}u'' + 2\frac{k}{\epsilon}\prt_xu' + \frac{\prt_xk}{\epsilon}u' + \prt_{xx}u\right] + \alpha \left(\frac{kq}{\epsilon}v' +\prt_yv\right) = 0,   \label{eq:uBO}
\ee
where again $f' \equiv \prt_{\theta}f$. Again the kinematic Eqs.~(\ref{eq:k}), (\ref{eq:q}), (\ref{eq:kl}) are the first Whitham equations for 2DBO. Substituting Eq.~(\ref{eq:v0}) into the leading order of Eq.~(\ref{eq:uBO}) implies
\be
-kVu_0' + ku_0u_0' + k^2\mathcal H[u_0''] = 0.   \label{eq:0BO}
\ee
Then integration leads to
$$
-Vu_0 + \frac{u_0^2}{2} + k\mathcal H[u_0'] = C_1.  
$$
Its physically relevant periodic solution can be written as
\be
u_0 = \frac{4k^2}{\sqrt{A^2+4k^2} - A\cos(\theta-\theta_*)} + \beta,   \label{eq:u0BO}
\ee
where the slow variables $A$, $k$, $\beta$ and $V$ satisfy the relation
\be
V = \frac{1}{2}\sqrt{A^2+4k^2} + \beta.   \label{eq:abmBO}
\ee
At the subsequent order in $\epsilon$ we get, after substituting Eq.~(\ref{eq:v0}) and the expression for $v_1'$ from Eq.~(\ref{eq:O1system2}) into Eq.~(\ref{eq:uBO})
\begin{multline}
k(k\mathcal H[u_1''] + ((u_0 - V) u_1)') + 
\\
+ \prt_tu_0 + u_0\prt_xu_0 + \mathcal H[2k\prt_xu_0' + \prt_xku_0'] + \alpha\left( q^2\prt_xu_0 + 2qD_yu_0 + u_0D_yq + D_yp \right) = 0.   \label{eq:u1BO}
\end{multline}
The other Whitham equations can be derived as secularity conditions ensuring that the solutions $u_1$ and $v_1$ of Eqs.~(\ref{eq:u1BO}), (\ref{eq:O1system2}) are periodic rather than growing in $\theta$. Imposing periodicity and integrating Eqs.~(\ref{eq:u1BO}) and (\ref{eq:O1system2}) over the period in $\theta$ leads to two secularity conditions, respectively,
\be
\prt_t\overline{u_0} + \frac{1}{2}\prt_x(\overline{u_0^2}) + \alpha\left( q^2\prt_x\overline{u_0} + 2qD_y\overline{u_0} + \overline{u_0}D_yq + D_yp \right) = 0,   \label{eq:S1BO}
\ee
and Eq.~(\ref{eq:S3}) which we rewrite below
\be
\prt_xp = D_yQ - Q\prt_xq.   \label{eq:S3BO}
\ee
We denote this equation as (\ref{eq:S3BO}) because the value of $Q$ comes from the 2DBO equation. The third (and the last needed) secularity condition is readily obtained when one notices that the terms depending on $u_1$ in Eq.~(\ref{eq:u1BO}) are the same as for BO (and $v_1$ is absent in it). So we multiply Eq.~(\ref{eq:u1BO}) by $u_0$ and integrate over the period to find
\be
\prt_t\overline{u_0^2} + \frac{2}{3}\prt_x\overline{u_0^3} + 2\overline{u_0\mathcal H[2k\prt_xu_0' + \prt_xku_0']} + \alpha\left( q^2\prt_x\overline{u_0^2} + 2D_y(q\overline{u_0^2}) + 2\overline{u_0}D_yp \right) = 0.  \label{eq:S2BO}
\ee
The averages over the period entering the secularity equations have the following expressions in terms of the parameters in Eq.~(\ref{eq:u0BO}) \cite{ABW-BO} 
(see also \cite{Mats98}):
\bse
\begin{gather}
Q \equiv \overline{u_0} = 2\pi(2k+\beta),   \label{eq:QBO}
\\
Q_2 \equiv \overline{u_0^2} = 2\pi(4Vk+\beta^2),   \label{eq:Q2BO}
\\
Q_3 \equiv \overline{u_0^3} = 2\pi(8k^3 + 3kA^2 + 12Vk\beta - 6k\beta^2 + \beta^3),   \label{eq:Q3BO}
\\
\prt_xG \equiv -2\overline{u_0\mathcal H[2k\prt_xu_0' + \prt_xku_0']} = 2\pi\prt_x(kA^2).   \label{eq:GBO}
\end{gather}
\ese
The secularity equations (\ref{eq:S1BO}),~(\ref{eq:S3BO}),~(\ref{eq:S2BO}) and the kinematic equations (\ref{eq:k}) and~(\ref{eq:q}) comprise the Whitham 2DBO equations in physical coordinates.

\subsection{Whitham equations in BO Riemann variables}

The original one-dimensional Benjamin-Ono (BO) equation is known~\cite{Mats98} to have Riemann variables $r_j$, $j=1,2,3$, with $r_1\le r_2\le r_3$, which are defined as follows:
\be
V = r_2+r_3,   \qquad  k=r_3-r_2,   \qquad  \beta = 2r_1.   \label{eq:rs}
\ee
The relation Eq.~(\ref{eq:abmBO}) implies that
\be
\sqrt{A^2+4k^2} = 2(r_2+r_3-2r_1),   \qquad   A = 4\sqrt{(r_2-r_1)(r_3-r_1)}.   \label{eq:Ar}
\ee
In terms of these Riemann variables the leading order solution $u_0$ is
\be
u_0 = \frac{2(r_3-r_2)^2}{r_2+r_3-2r_1 - 2\sqrt{(r_2-r_1)(r_3-r_1)}\cos(\theta-\theta_*)} + 2r_1.   \label{eq:u0r}
\ee
Then, in terms of the $r$-variables, the averages in the Whitham equations read
\bse
\label{eq:QrBO}
\begin{gather}
\frac{Q}{2\pi} = 2(r_3-r_2+r_1),
\\
\frac{Q_2}{2\pi} = 4(r_3^2-r_2^2+r_1^2),   
\\
\frac{Q_3}{2\pi} = 8(r_3^3-r_2^3+r_1^3) + 24(r_2-r_1)(r_3-r_2)(r_3-r_1),
\\
\frac{G}{2\pi} = kA^2 = 16(r_2-r_1)(r_3-r_2)(r_3-r_1).   
\end{gather}
\ese
Similar to the KP-Whitham system, only three equations, namely~(\ref{eq:k}), (\ref{eq:S1BO}) and~(\ref{eq:S2BO}), are further substantially transformed. 
After substituting these expressions into Eqs.~(\ref{eq:k}), (\ref{eq:S1BO}), (\ref{eq:S2BO}) the latter are brought into the form, 
\bse
\begin{gather}
\prt_t(r_3-r_2) + \prt_x(r_3^2-r_2^2) + \alpha\prt_x(q^2(r_3-r_2)) = 0,   
\label{eq:kr}
\\
\prt_t(r_3-r_2+r_1) + \prt_x(r_3^2-r_2^2+r_1^2) + \frac{\alpha}{2}(D_y+q\prt_x)(2q(r_3-r_2+r_1) + p) = 0,   
\label{eq:S1r}
\\
\prt_t(r_3^2-r_2^2+r_1^2) + \frac{4}{3}\prt_x(r_3^3-r_2^3+r_1^3) + 
\kern18em{ }\nonumber
\\
\kern4em{ }
+\frac{\alpha}{4}(4q^2\prt_x(r_3^2-r_2^2+r_1^2) + 8D_y(q(r_3^2-r_2^2+r_1^2)) + 4(r_3-r_2+r_1)D_yp) = 0,  
\label{eq:S2r}
\end{gather}
\ese
respectively.
Now, following what we did for the KP equation, we transform the last three equations to the form that would be diagonal in the $r$-variables if $\alpha$ were zero. Let
%
\begin{multline}
Y_0 = \prt_x(q^2(r_3-r_2)),   \qquad  Y_1 = (D_y+q\prt_x)(2q(r_3-r_2+r_1) + p),  
\\
Y_2 = 4q^2\prt_x(r_3^2-r_2^2+r_1^2) + 8D_y(q(r_3^2-r_2^2+r_1^2)) + 4(r_3-r_2+r_1)D_yp.  \label{eq:YsBO}
\end{multline}
%
Subtracting Eq.~(\ref{eq:kr}) from Eq.~(\ref{eq:S1r}), we get
\be
\prt_tr_1 + 2r_1\prt_xr_1 + \alpha(Y_1/2 - Y_0) = 0.   \label{eq:0r1}   
\ee
Taking the combination of equations Eq.~(\ref{eq:S2r}) minus $(r_3+r_2)$ times Eq.~(\ref{eq:kr}) minus $2r_1$ times (\ref{eq:0r1}) yields
\bse
\be
\prt_t(r_3+r_2) + 2(r_3\prt_xr_3 + r_2\prt_xr_2) + \alpha\frac{(Y_2/4 - r_1Y_1 + (2r_1-r_2-r_3)Y_0)}{(r_3-r_2)} = 0.   \label{eq:Vr}
\ee
Adding and subtracting Eqs.~(\ref{eq:Vr}), (\ref{eq:kr}) yields, respectively,
\be
\prt_tr_3 + 2r_3\prt_xr_3 + \alpha\frac{(Y_2/4 - r_1Y_1 - 2(r_2-r_1)Y_0)}{2(r_3-r_2)} = 0,   \label{eq:0r3}
\ee
\be
\prt_tr_2 + 2r_2\prt_xr_2 + \alpha\frac{(Y_2/4 - r_1Y_1 - 2(r_3-r_1)Y_0)}{2(r_3-r_2)} = 0.   \label{eq:0r2}
\ee
\ese
To obtain the final form of the Whitham Eqs.~(\ref{eq:r1}), $j=1,2,3$, we use the expressions Eq.~(\ref{eq:YsBO}) together with Eq.~(\ref{eq:kl}) rewritten for the present case as
\be
(r_3-r_2)\prt_xq = D_y(r_3-r_2),  \label{eq:kq}
\ee
and Eq.~(\ref{eq:S3BO}) which now takes form
\be
\prt_xp = 2D_y(r_3-r_2+r_1) - 2(r_3-r_2+r_1)\prt_xq.   
\label{eq:S3r}
\ee
Taking Eq.~(\ref{eq:kq}) into account, Eq.~\eqref{eq:S3r} simplifies to Eq.~(\ref{eq:pBO}) of section 1. 
Then, using Eqs.~(\ref{eq:YsBO}), (\ref{eq:kq}) and~(\ref{eq:pBO}), one can show that Eqs.~(\ref{eq:0r1}), (\ref{eq:0r2}) and~(\ref{eq:0r3}) can be written in the form of Eq.~(\ref{eq:r1}) in section~1.
These equations constitute the final 2DBO-Whitham system, together with Eq.~(\ref{eq:pBO}) just mentioned and Eq.~(\ref{eq:qBO}) (which comes from equation (\ref{eq:q}) with the above identification of $V$).

This result agrees with that obtained in \cite{ABW-BO}. Like with the KP-Whitham system, however, using this approach does not introduce (removable) singularities.

\section{Derivation of the m2KP Whitham system}
\label{s:m2KP}

\subsection{Modulation equations}

The {\it defocusing} m2KP equation can be written as the system,
\be
\prt_tu - 6u^2\prt_xu + \epsilon^2\prt_{xxx}u + \alpha \prt_yv = 0,   \label{eq:mu0}
\ee
and Eq.~(\ref{eq:v}). Then, after introducing fast and slow scales, Eq.~(\ref{eq:mu0}) takes form
\be
\left(-\frac{\omega}{\epsilon}\prt_{\theta} + \prt_t\right)u - 6u^2\left(\frac{k}{\epsilon}\prt_{\theta} + \prt_x\right)u + \epsilon^2\left(\frac{k}{\epsilon}\prt_{\theta} + \prt_x\right)^3u + \alpha \left(\frac{kq}{\epsilon}\prt_{\theta} +\prt_y\right)v = 0,   \label{eq:mu}
\ee
Once again, the kinematic Eqs.~(\ref{eq:k}), (\ref{eq:q}), (\ref{eq:kl}) are basic Whitham equations for m2KP. After substituting  Eq.~(\ref{eq:v0}) into the leading order of Eq.~(\ref{eq:mu}) we find 
\be
-kVu_0' - 6ku_0^2u_0' + k^3u_0''' = 0,   \label{eq:0MK}
\ee
and, integrating it once, get
\be
-Vu_0 - 2u_0^3 + k^2u_0'' = A_1.  \label{eq:I1MK}
\ee
Multiplying Eq.~(\ref{eq:I1MK}) by $2u_0'$ and integrating again, one obtains the equation for elliptic (``genus one'') solution of m2KP (or mKdV),
\be
k^2(u_0')^2 = u_0^4 + Vu_0^2 + 2A_1u_0 + A_2,   \label{eq:g1a} 
\ee
or 
\be
k^2(u_0')^2 = (u_0-a)(u_0-b)(u_0-c)(u_0-d),  \qquad   a\ge b\ge c\ge d.    \label{eq:g1b} 
\ee
A bounded periodic solution can be written in terms of Jacobian elliptic functions as
\be
u_0 = a - \frac{a-b}{1 - \gamma\;\text{sn}^2\left(2K(m)(\theta - \theta_*); m\right)},   \label{eq:u0MK}
\ee
where $\theta_*$ is an arbitrary integration constant, 
\be
\gamma = \frac{b-c}{a-c},  \qquad  m = \frac{(b-c)(a-d)}{(a-c)(b-d)},   \label{eq:abmMK}
\ee
and we have the following relations among the parameters (slow variables):
\bse
\begin{gather}
a+b+c+d = 0,   \label{eq:sum}
\\
V = ab+bc+ca + (a+b+c)d = ab+bc+ca - (a+b+c)^2,  \label{eq:V}
\\
2A_1 = -abc - (ab+bc+ca)d = (a+b+c)(ab+bc+ca) - abc,  \label{eq:A1}
\\
A_2 = abcd = -(a+b+c)abc.  \label{eq:A2}
\end{gather}
\ese
Note that, for the case of real roots $a,b,c,d$, the phase velocity $V$ must be {\it negative}; so the waves travel only in one direction. The normalization of the elliptic function with fixed period one implies that
\be
k^2 = \frac{(a-c)(b-d)}{16K^2(m)},   \label{eq:mk2}
\ee
where $K(m)$ is the first complete elliptic integral. Then $c\le u_0 \le b$, i.e.~$u_0$ oscillates between the two middle roots of the quartic in Eq.~(\ref{eq:g1b}).

At the next order in $\epsilon$ we get, after substituting Eq.~(\ref{eq:v0}) and the expression for $v_1'$ from Eq.~(\ref{eq:O1system2}) into Eq.~(\ref{eq:mu}),
\begin{multline}
k(k^2u_1'' - 6u_0^2u_1 - V u_1)' +     
\\
+ \prt_tu_0 - 6u_0^2\prt_xu_0 + 3k^2\prt_xu_0'' + 3k\prt_xku_0'' + \alpha\left( q^2\prt_xu_0 + 2qD_yu_0 + u_0D_yq + D_yp \right) = 0.   \label{eq:mu1}
\end{multline}
Similar to the KP and 2DBO equations, the other Whitham equations can be derived as secularity conditions ensuring that the solutions $u_1$ and $v_1$ of Eqs.~(\ref{eq:mu1}), (\ref{eq:O1system2}) are periodic rather than growing in $\theta$. Imposing periodicity and integrating Eqs.~(\ref{eq:mu1}) and (\ref{eq:O1system2}) over the period in $\theta$ leads to two secularity conditions, respectively,
\be
\prt_t\overline{u_0} - 2\prt_x(\overline{u_0^2}) + \alpha\left( q^2\prt_x\overline{u_0} + 2qD_y\overline{u_0} + \overline{u_0}D_yq + D_yp \right) = 0,   \label{eq:S1MK0}
\ee
and Eq.~(\ref{eq:S3}). The third (and the last needed) secularity condition is readily obtained when one notices that the terms depending on $u_1$ in Eq.~(\ref{eq:mu1}) are the same as for mKdV (and $v_1$ is absent in it). 
So we multiply Eq.~(\ref{eq:mu1}) by $u_0$ and integrate over the period to find
\be
\prt_t\overline{u_0^2} - 3\prt_x\overline{u_0^4} - 3\prt_x(k^2\overline{(u_0')^2}) + \alpha\left( q^2\prt_x\overline{u_0^2} + 2D_y(q\overline{u_0^2}) + 2\overline{u_0}D_yp \right) = 0.  \label{eq:S2MK0}
\ee
We also use the notation Eq.~(\ref{eq:Qn}); from Eq.~(\ref{eq:u0MK}) we obtain
\bse
\begin{gather}
Q \equiv \overline{u_0} = a - (a-b)\frac{\Pi}{K},   \label{eq:QMK}
\\
Q_2 \equiv \overline{u_0^2} = \frac{a^2+a(b+c)-bc}{2} - \frac{(a-c)(a+2b+c)}{2}\frac{E}{K},   \label{eq:Q2MK}
\end{gather}
\ese
where $K=K(m)$, $E=E(m)$ and $\Pi = \Pi(\gamma,m)$ are the first, second and third complete elliptic integrals, respectively. 
Integrating Eq.~(\ref{eq:I1MK}) we find
\be
-2Q_3 \equiv -2\overline{u_0^3} = VQ+A_1.   \label{eq:Q3MK}
\ee
Integration of Eq.~(\ref{eq:g1a}) gives 
\be
G \equiv k^2\overline{(u_0')^2} = Q_4 + VQ_2 + 2A_1Q + A_2,   \label{eq:mIg1}
\ee
while multiplying Eq.~(\ref{eq:I1MK}) by $u_0$ and integrating gives 
\be
-G = 2Q_4 + VQ_2 + A_1Q.   \label{eq:mIg2}
\ee
Combining Eqs.~(\ref{eq:mIg1}), (\ref{eq:mIg2}) yields 
\bse
\begin{gather}
3Q_4 = -2VQ_2 - 3A_1Q - A_2,   \label{eq:Q4}
\\
3G = VQ_2 + 3A_1Q + 2A_2.   \label{eq:GMK}
\end{gather}
\ese
Using Eqs.~(\ref{eq:Q3MK}), (\ref{eq:Q4}), (\ref{eq:GMK}), we get the secularity equations in the form
\bse
\begin{gather}
\prt_tQ + \prt_x(VQ+A_1) + \alpha(D_y+q\prt_x)(qQ+p) = 0,   \label{eq:S1MK}
\\
\prt_tQ_2 + \prt_x\left(VQ_2-A_2\right) + \alpha\left[ q^2\prt_xQ_2 + 2D_y(qQ_2) + 2QD_yp \right] = 0,  \label{eq:S2MK}
\\
\prt_xp = D_yQ - Q\prt_xq.   \label{eq:S3MK}
\end{gather}
\ese
The secularity equations (\ref{eq:S1MK}), (\ref{eq:S2MK}) and~(\ref{eq:S3MK}) and the kinematic equations (\ref{eq:k}) and~(\ref{eq:q}) comprise the m2KP-Whitham equations in physical coordinates.

\subsection{Whitham equations in mKdV Riemann variables}

The three independent roots of the quartic in Eq.~(\ref{eq:g1b}), which we label $a,b,c$, are simply related to the mKdV Riemann variables by~\cite{DN-mKdV}
\be
a = r_2 + r_3 - r_1,    \qquad  b = r_3 + r_1 - r_2,   \qquad  c = r_1 + r_2 - r_3,   \label{eq:la-R}
\ee
i.e.~exactly as in the KdV case (see section on KP equation). From Eq.~(\ref{eq:la-R}), we express the functions of the roots $V$, $A_1$ and $A_2$ in terms of the $r$-variables,
\be
V = -2(r_1^2+r_2^2+r_3^2),   \qquad  A_1 = 4r_1r_2r_3,   \qquad  A_2 = r_1^4+r_2^4+r_3^4 - 2(r_1^2r_2^2+r_2^2r_3^2+r_3^2r_1^2).    \label{eq:e-R}
\ee
We want to express the Eqs.~(\ref{eq:k}), (\ref{eq:q}), (\ref{eq:kl}), (\ref{eq:S1MK}), (\ref{eq:S2MK}), (\ref{eq:S3MK}) in terms of Riemann variables for mKdV, $r_i$, $i=1,2,3$. With the ``total" time derivative $D = \prt_t + V\prt_x$, we rewrite Eqs.~(\ref{eq:k}), (\ref{eq:S1MK}) and~(\ref{eq:S2MK}) respectively as
\bse
\begin{gather}
\frac{Dk}{k} + \prt_xV + \alpha Y_0 = 0,   \label{eq:mk}
\\
DQ + Q\prt_xV + \prt_xA_1 + \alpha Y_1 = 0,  \label{eq:mQ}
\\
DQ_2 + Q_2\prt_xV - \prt_xA_2 + \alpha Y_2 = 0,   \label{eq:mQ2}
\end{gather}
\ese
where we denoted for further use
%
\begin{multline}
Y_0 = \frac{\prt_x(kq^2)}{k},  \qquad Y_1 = (D_y+q\prt_x)(qQ+p) = q^2\prt_xQ + 2qD_yQ + QD_yq + D_yp, 
\\
Y_2 = q^2\prt_xQ_2 + 2D_y(qQ_2) + 2QD_yp.   \label{eq:YsMK}
\end{multline}
%
Similar to KP and 2DBO, using above Riemann $r$-variables we transform Eqs.~(\ref{eq:mk}), (\ref{eq:mQ}), (\ref{eq:mQ2}) and then we diagonalize the mKdV parts of Eqs.~(\ref{eq:mk}), (\ref{eq:mQ}), (\ref{eq:mQ2}). The details are given in Appendix C. This yields three of the Whitham equations in the form given in section 1. One can readily verify, using Eqs.~(\ref{eq:QR}), (\ref{eq:mgR}), that the last terms in Eqs.~(\ref{eq:Rj}) remain finite in the limits as $m\to 0$ and $m\to 1$ when their denominators become zero due to the limits $r_2^2\to r_3^2$ or $r_2^2\to r_1^2$, respectively. In the other Whitham equations, Eqs.~(\ref{eq:q}), (\ref{eq:S3MK}), (\ref{eq:kl}), we use Eqs.~(\ref{eq:e-R}) to bring them to the final form in terms of $r_j$, $q$ and $p$-variables.

\section{Stability analysis of the periodic solutions of the m2KP equation}
\label{s:m2KPstability}

The leading order solution~\eqref{eq:u0MK} in terms of the Riemann-type variables~$r_1,r_2,r_3$ is
\be
u_0(x,t) = r_2+r_3-r_1 - \frac{2(r_2-r_1)}{1 - \gamma\;\text{sn}^2\left(2K(m)(\theta - \theta_*); m\right)},   
\label{eq:u0R}
\ee
where
\vspace*{-1ex}
\be
\gamma = \frac{r_3-r_2}{r_3-r_1},  \qquad  m = \frac{r_3^2-r_2^2}{r_3^2-r_1^2}.   \label{eq:abmR}
\ee
\unskip
Exact periodic solutions of the m2KP equation have the form Eq.~(\ref{eq:u0R}) with constant values of $r_1,r_2,r_3,q$ and $p$. 
In other words, the periodic solutions of the m2KP equation correspond to constant solutions of the m2KP-Whitham equations.
As a special case, when $q=0$, Eq.~\eqref{eq:u0R} yields the cnoidal soluions of the mKdV equation.
Like for the KdV and BO equations, DSW of the mKdV equation can be described as a slow modulation of the cnoidal waves.
One such solution is shown in Fig.~\ref{f:m2KP}(left), and is described in detail in Appendix~D. 

\begin{figure}[t!]
\centerline{\qquad\raise8ex\hbox{\includegraphics[width=6cm]{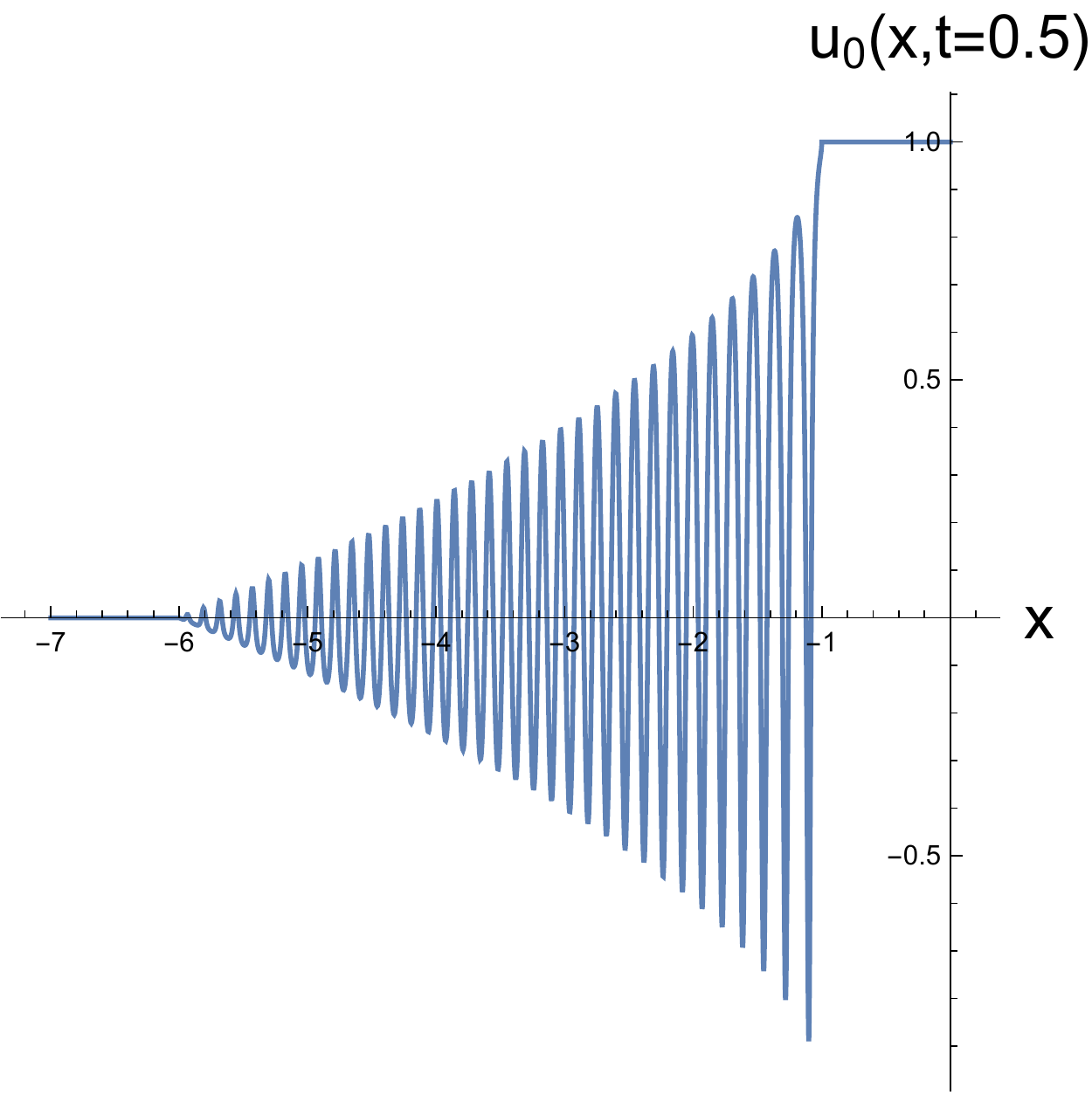}}\qquad
\includegraphics[width=7.85cm]{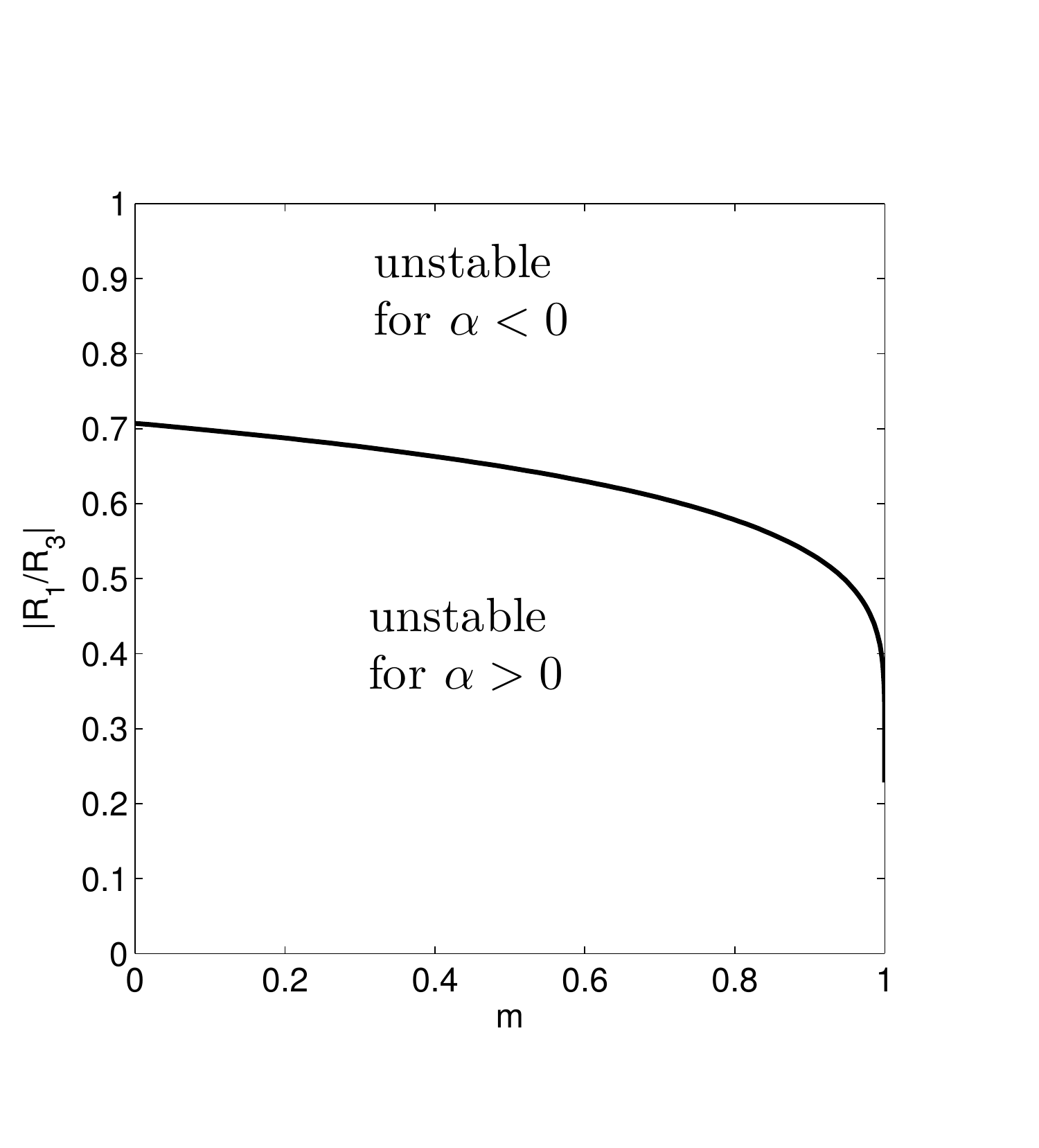}}
\vglue-3\medskipamount
\caption{Left: A typical DSW solution of the mKdV equation with step IC for $\epsilon=0.05$ and $t=0.5$ using Eqs.~\eqref{eq:u0R} and~\eqref{eq:mGP}
(cf.\ Appendix D for details).
Right: Stability-instability regions of the periodic solutions of the m2KP equation (see text for details).} 
\label{f:m2KP}
\vskip1.4\bigskipamount
\centerline{%
\includegraphics[width=5cm]{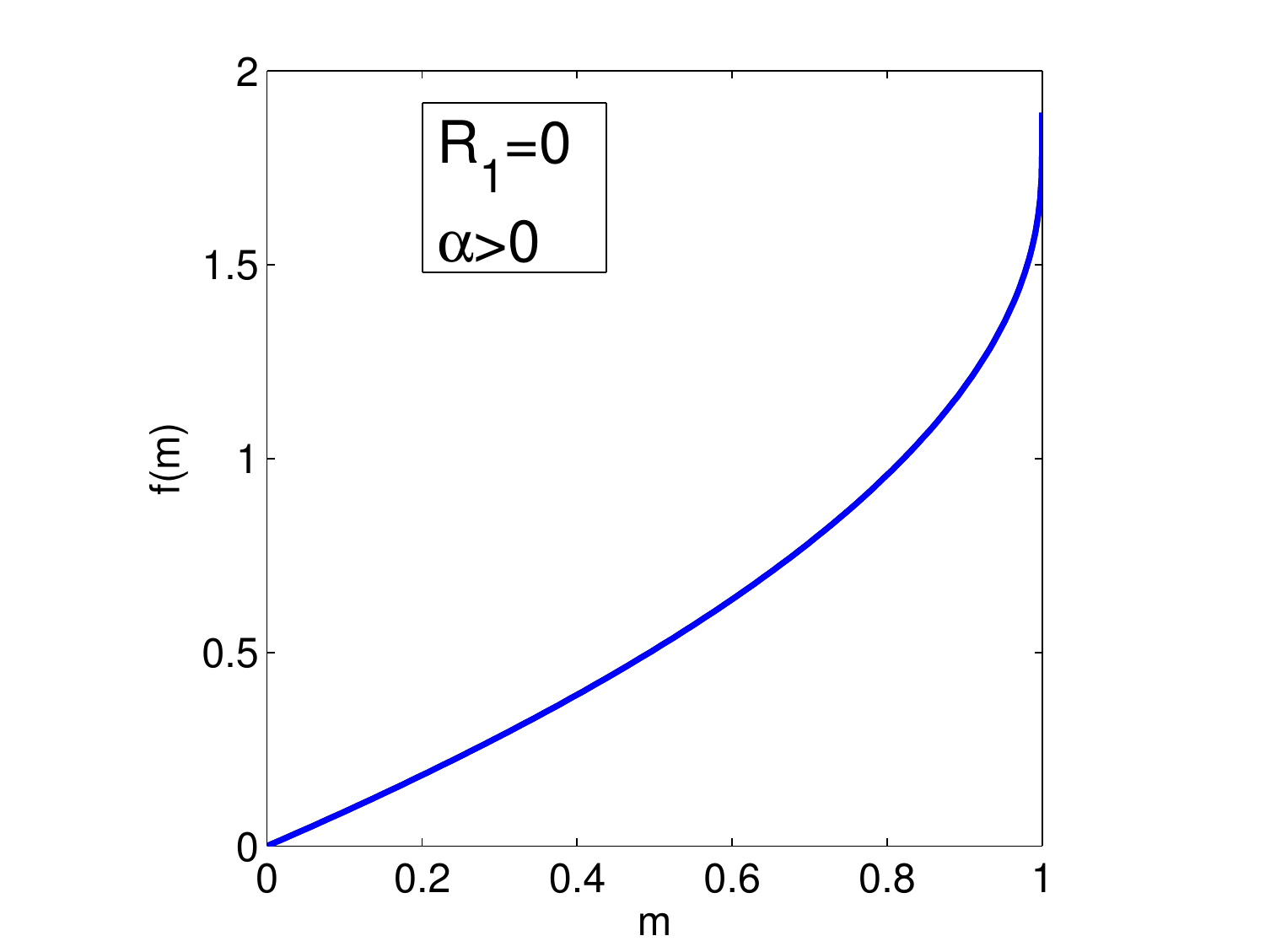}\hglue-2em
\includegraphics[width=5cm]{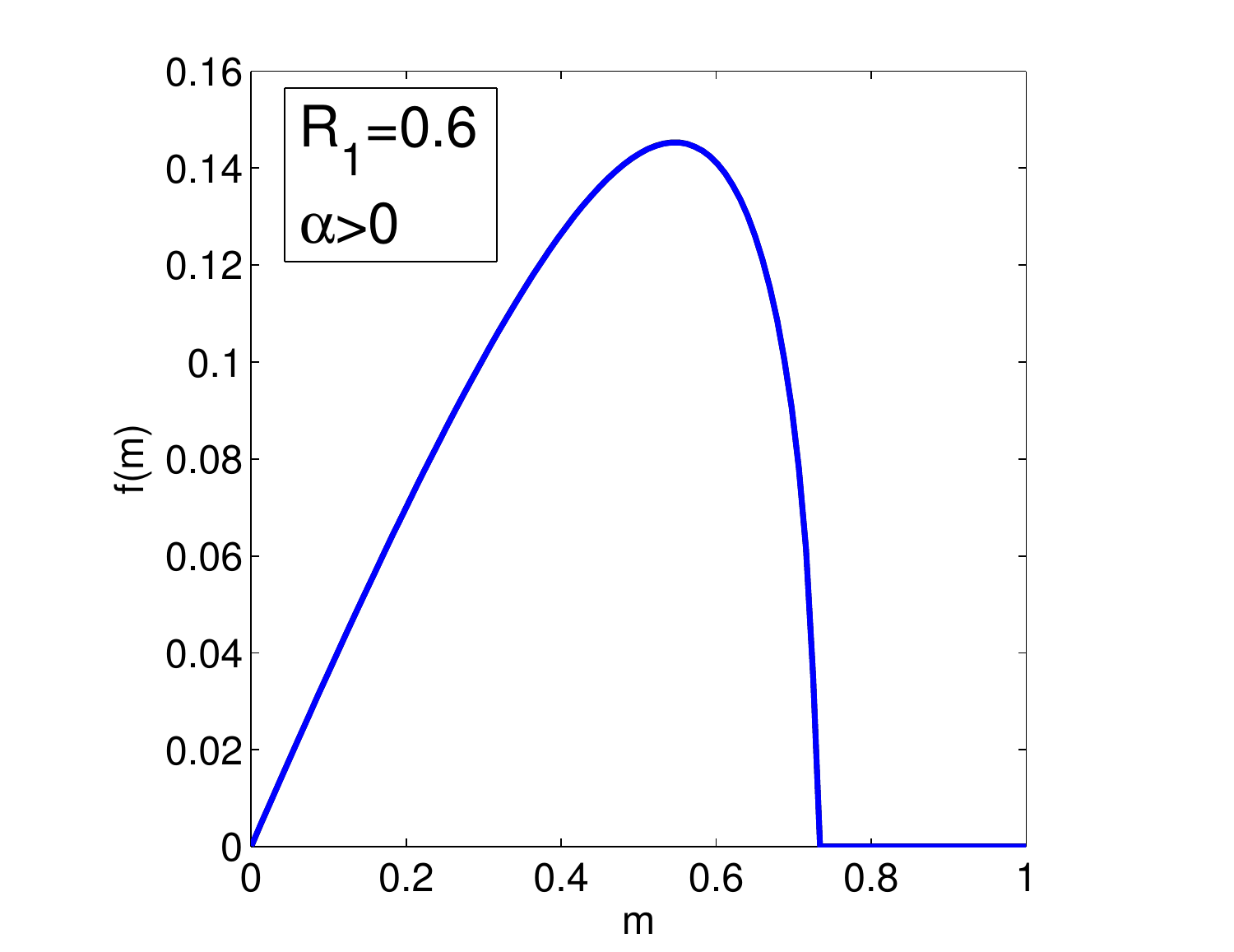}\hglue-2em
\includegraphics[width=5cm]{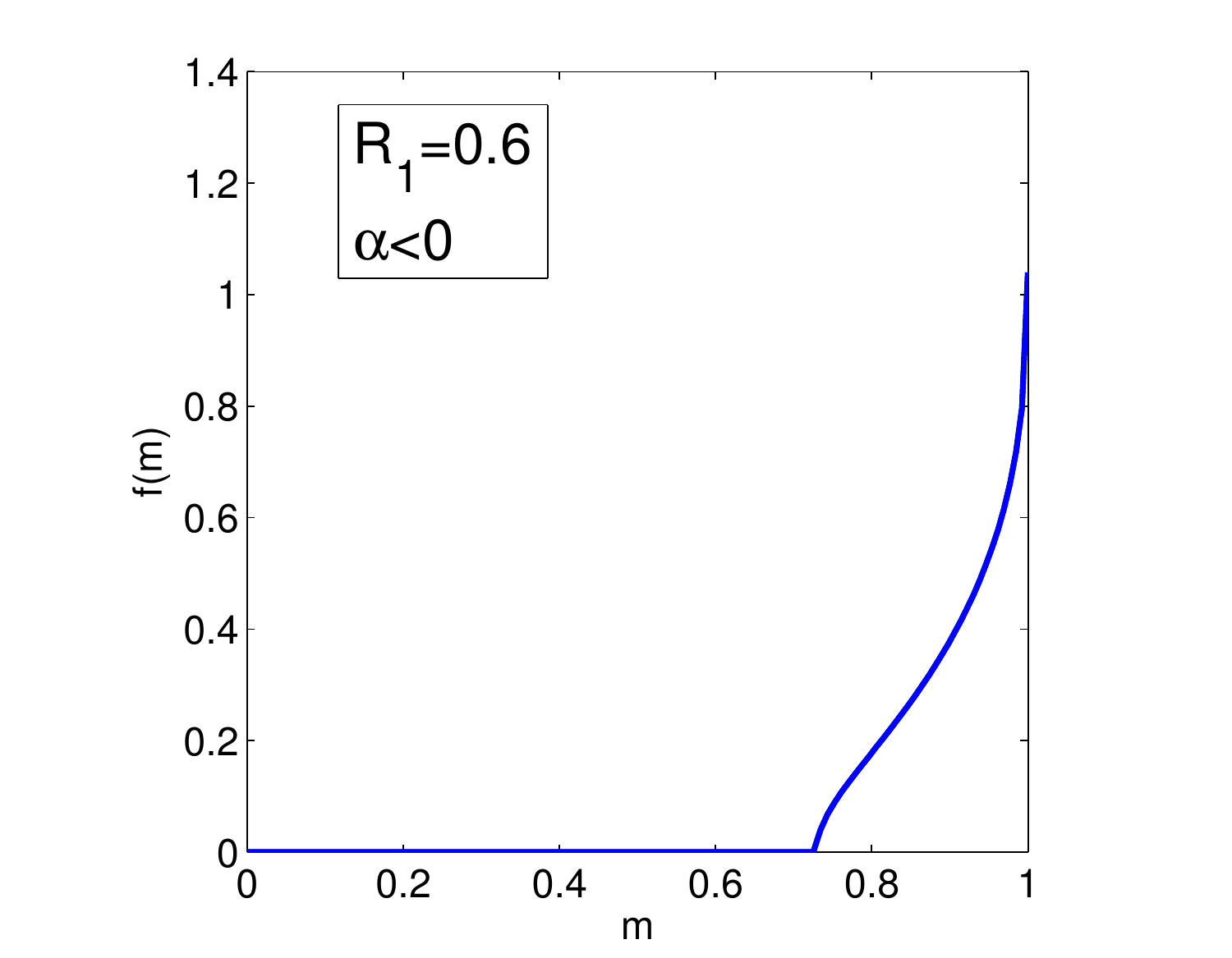}\hglue-2em
\includegraphics[width=5cm]{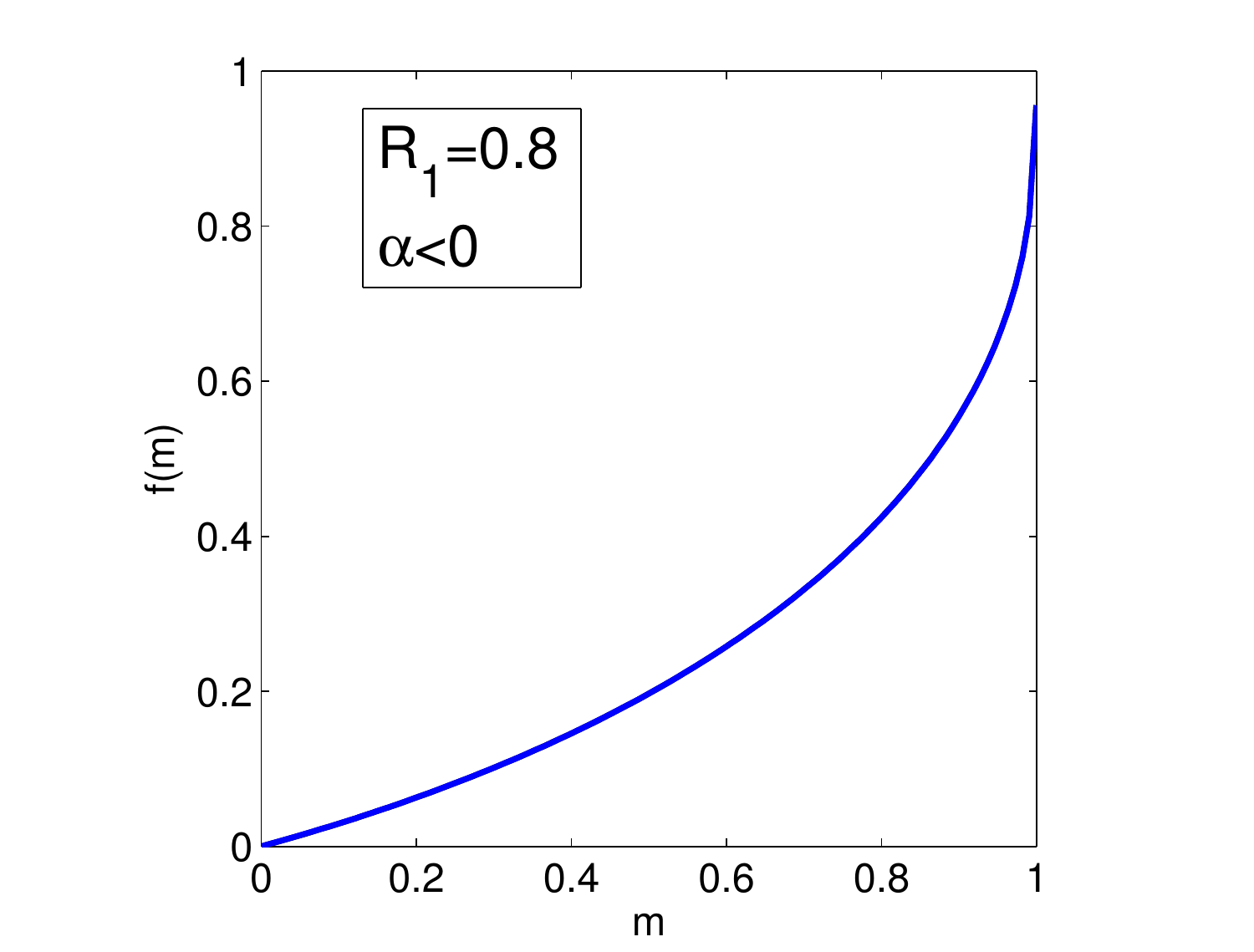}%
}
\caption{ Instability growth rates relative to the transverse wavenumber for periodic solutions of m2KP equation. From left to right: cases $r_1=0,\alpha>0$; $r_1=0.6,\alpha>0$; $r_1=0.6,\alpha<0$ and $r_1=0.8,\alpha<0$. In each plot both theoretical and numerical rates are presented but they lie on top of one another, hence they are difficult to distinguish from each other. } 
\end{figure}

Next we use the m2KP-Whitham system to study stability of the cnoidal wave solutions of the m2KP equation.
We do so by considering small perturbations to the periodic solution~\eqref{eq:u0R} with constant parameters.
That is, we take $r_j = \bar r_j+\delta r_j$, $q=\delta q$, $p=\delta p$, with $\bar r_j$ constant for $j=1,2,3$, and 
where for simplicity we set $\bar q = \bar p = 0$,
and we linearize the Whitham equations by taking $\delta r_j$, $\delta q$ and $\delta p$ to be small. 
We consider plane wave perturbations of the form 

\be
\delta r_j = \rho_j\, e^{i(\kappa x + ly - \omega t)},\quad j=1,2,3,
\qquad
\delta q = \eta\, e^{i(\kappa x + ly - \omega t)},\qquad 
\delta p =  \upsilon\, e^{i(\kappa x + ly - \omega t)}\,. 
\ee
\unskip
The five linearized Whitham equations then reduce to the following homogeneous linear algebraic system of equations:
\vspace*{-1ex}
\bse
\begin{gather}
(\kappa V_j - \omega)\rho_j + \alpha l(\Phi_j\eta + \Psi_j\upsilon) = 0,  \qquad j=1,2,3, 
\label{eq:roj}
\\
(\kappa V - \omega)\eta - 4l\sum_j\bar r_j\rho_j = 0,  
\label{eq:ql}
\\
\kappa \upsilon - l\sum_j\Pi_j\rho_j = 0,   
\label{eq:pl}
\end{gather}
\ese
where the coefficients are $V_j = v_j(\bar r)$, $V = V(\bar r)$, 
\begin{gather*}
\Phi_j = \frac{\bar r_j(\bar r_jQ_2-\bar r_i\bar r_lQ)}{\frac{\prt_jk}{k}(\bar r_j^2-\bar r_l^2)(\bar r_j^2-\bar r_l^2)},   \quad  
\Psi_j = \frac{\bar r_j(\bar r_jQ-\bar r_i\bar r_l)}{\frac{\prt_jk}{k}(\bar r_j^2-\bar r_l^2)(\bar r_j^2-\bar r_l^2)},\quad
\Pi_j = -\frac{\bar r_i\bar r_l}{\bar r_j}\frac{\prt_jk}{k},\quad  j\neq i \neq l\neq j,
\\
\noalign{\noindent and}
Q=Q(\bar r),\quad Q_2 = Q_2(\bar r),\quad \prt_jQ = \prt_jQ(\bar r),\quad  \prt_jk/k = (\prt_jk/k)(\bar r)
\end{gather*}
[see Eqs.~\eqref{e:m2KPcoeffs} in section~\ref{s:main} for the definitions of $Q$, $Q_2$ and $\prt_j k/k$].
Equating the determinant of the system to zero yields the stability/dispersion relation for the perturbation parameters $\omega$, $\kappa$ and $l$. In general this relation is somewhat complicated but it simplifies in two important special cases. For longitudinal perturbations $l=0$ the relation reduces to
 
$$
(\omega-V\kappa)\prod_j(\omega-V_j\kappa)=0,
$$
which shows that the solution is linearly stable with respect to such perturbations since all eigenvalues $\omega$ are real for real $\kappa$. This is expected since the problem has been reduced to one dimension; i.e. defocusing mKdV which is known to be stable.
 
The interesting case, however, is $\kappa=0$. This corresponds to purely transverse perturbations. Since we are within Whitham theory these perturbations are relatively slow; i.e.~they are long wave perturbations. Then the dispersion relation reduces to the form $\omega(\omega^2+\alpha l^2f(r_1,r_2,r_3))=0$. Besides the simple real solution $\omega=0$, its eigenfrequencies are given by the following explicit formula (here and further on we write $r_j$ instead of $\bar r_j$ but the constant values are implied):
%
%
$$
\omega^2/l^2 = -4\alpha \frac{(Q_2-Q^2)[r_3^2\frac{E}{K}(\frac{E}{K}-1+m) - r_2^2\frac{E}{K}(1-\frac{E}{K}) - r_1^2(1-\frac{E}{K})(\frac{E}{K}-1+m)]}{(r_3^2-r_1^2)\frac{E}{K}(1-\frac{E}{K})(\frac{E}{K}-1+m)}.
$$
Here all factors multiplying $-\alpha$ on the right-hand side are nonnegative except for the last factor in the numerator which can change sign. This last factor can be rewritten as
\be
h(r_1,r_2,r_3) = r_3^2\left[\frac{E}{K}\left((2-m)\frac{E}{K} - 2(1-m)\right) - \frac{r_1^2}{r_3^2}\left(1-\frac{E}{K}\right)\left((1+m)\frac{E}{K} - 1 + m\right)\right],   \label{eq:St}
\ee
which shows that the stability essentially depends on two parameters, $m$ and $(r_1/r_3)^2$. For $r_1=0$, which we took in Appendix D to describe a DSW solution, the factor $h$ can be shown to be nonnegative for all $m$. Thus, in this case the periodic solution Eq.~(\ref{eq:u0R}) is linearly unstable for $\alpha>0$ and stable for $\alpha<0$. This is exactly opposite to the stability dependence on $\alpha$ for KP~\cite{ABW-KP}. However, as $(r_1/r_3)^2$ increases from zero, one sees from Eq.~(\ref{eq:St}) that $h$ decreases for any fixed $m$, $0<m<1$. This in fact leads to the change of sign of $h$ on a certain curve in the domain of parameters $0\le m\le 1$, $0\le (r_1/r_3)^2 < 1$, given by the equation $h(m,(r_1/r_3)^2)=0$, see Fig.~1(right). Since $h$ is always zero for $m=0$ or $m=1$, the curve where the sign changes (and therefore stability changes) ends at certain points on the intervals $m=0$ and $m=1$. Thus, e.g.~for the cases when $r_3^2-r_1^2\ll r_3^2$ the periodic solution is stable for $\alpha>0$ and unstable for $\alpha<0$, opposite to the case $r_1=0$. The left end (at $m=0$) of the stability boundary curve occurs at $|r_1/r_3|=1/\sqrt2 \approx 0.7071$. It turns out, however, that its right end occurs at the corner $m=1, r_1=0$ of the parameter $m, |r_1/r_3|$-plane. This means that the complete stability for $\alpha<0$ can be reached only when $r_1=0$. All the above results agree well with numerical computations of linear spectral stability, see Figs.~1(right) and 2.

Thus, for $\alpha<0$ (like in KP I), a stable DSW may exist but only if $r_1\equiv 0$. On the other hand, for $\alpha>0$ (like in KP II), a stable DSW may only exist if the ratio $r_1^2/r_3^2 > 0.5$. In all other cases the DSWs are unstable for some range of $m$.

\section{Unified derivation of (2+1)-dimensional Whitham equations for KP-type systems}
\label{s:derivation}

As discussed in section~\ref{s:main}, two of the modulation equations are an immediate consequence of the definition of the fast variable $\theta$.
In this section we show in general how the remaining three equations arise as secularity conditions in the multiple scales expansion for all equations of the form~\eqref{eq:1.1}.

Following the definition of $\theta$ via Eqs.~\eqref{eq:klo}, we substitute the expansion~\eqref{e:expansion} into the system~\eqref{e:evolutionsystem},
recalling that $\partial_x\ \to (k/\epsilon)\partial_\theta + \partial_x$, $\partial_y \to (l/\epsilon)\partial_\theta + \partial_y$ and $\partial_t \to -(\omega/\epsilon)\partial_\theta + \partial_t$.
We then expand the resulting equations in powers of~$\epsilon$.
In particular,
we expand $F(\cdot)$ as
\be
F(u, \partial_xu, \dots, \partial_x^nu; \epsilon) = 
\frac1\epsilon F^{(-1)}(u_0) + F_0^{(0)}(u_0) + F_1^{(0)}(u_0,u_1) + O(\epsilon)\,.
\label{e:Fexpansion}
\ee
(The fact that the leading-order term is at order $1/\epsilon$ is what allows one to obtain a nontrivial equation at
leading order.
In practice, the explicit dependence of $F(\cdot)$ on $\epsilon$ is usually determined precisely to ensure that this condition is satisfied.)\,
For example, denoting $\partial_{\theta}f = f'$ for brevity, for the KdV equation one has:
\vspace*{-0.2ex}
\bse
\begin{gather}
F^{(-1)}(u_0) = k^3u_0''' + 6k u_0u_0',\qquad
F_0^{(0)}(u_0) = 3k^2\prt_xu_0'' + 3k\prt_xku_0'' + 6u_0\prt_xu_0,\\
F_1^{(0)}(u_0,u_1) = k^3u_1''' + 6k(u_0u_1)'. 
\end{gather}
\ese
For the BO equation one has:
\bse
\begin{gather}
F^{(-1)}(u_0) = k^2H[u_0''] + ku_0u_0',\qquad
F_0^{(0)}(u_0) = 2kH[\prt_xu_0'] + \prt_xkH[u_0'] + ku_0\prt_xu_0,\\
F_1^{(0)}(u_0,u_1) = k^2H[u_1''] + k(u_0u_1)',
\end{gather}
\ese
Finally, for the mKdV equation one has:
\bse
\begin{gather}
F^{(-1)}(u_0) = k^3u_0''' - 6k u_0^2u_0',\qquad
F_0^{(0)}(u_0) = 3k^2\prt_xu_0'' + 3k\prt_xku_0'' - 6u_0^2\prt_xu_0,\\
F_1^{(0)}(u_0,u_1) = k^3u_1''' - 6k(u_0^2u_1)'. 
\end{gather}
\ese

We substitute Eqs.~\eqref{e:expansion} and~\eqref{e:Fexpansion} in the general system~\eqref{e:evolutionsystem}.
At leading order [i.e., $O(1/\epsilon)$] we have
\vspace*{-1ex}
\bse
\label{e:leadingordersystem}
\begin{gather}
- \omega u_0' + F^{(-1)}(u_0) + \alpha q k v_0' = 0\,,
\label{e:leadingordersystem1}
\\
kv_0' - qk u_0' = 0\,.
\label{e:leadingordersystem2}
\end{gather}
\ese
Equation~\eqref{e:leadingordersystem2} is readily solved to obtain Eq.~\eqref{eq:v0}. 
%
%
Substituting Eq.~\eqref{eq:v0} into the first one of Eqs.~\eqref{e:leadingordersystem} yields
\be
F^{(-1)}(u_0) - \Omega u_0' = 0\,,
\label{e:ODEu0}
\ee
where $\Omega = Vk$ and $V$ is given by Eq.~\eqref{e:qVdef} as before.
The solution of this ordinary differential equation (ODE) yields $u_0$ as a function of $\theta$,
in which, like with Eq.~\eqref{eq:v0}, all integration ``constants'' are actually functions of $(x,y,t)$,
to be determined at the next order.

At the next order in the expansion [i.e., $O(1)$] we have 
\vspace*{-1ex}
\bse
\label{e:O1system}
\begin{gather}
- \omega u_1' + F_1^{(0)}(u_0,u_1) + \alpha q k v_1' = -\prt_tu_0 - F_0^{(0)}(u_0) - \alpha\,\prt_y(qu_0+p)\,,
\label{e:O1system1}
\\
kv_1' - qk u_1' = \prt_yu_0 - \prt_x(qu_0 + p)\,.
\label{e:O1system2}
\end{gather}
\ese
One must now impose suitable conditions to prevent secular growth of the higher-order corrections $u_1$ and $v_1$.
In particular, three conditions must be imposed:
(i) zero-mean condition for the right-hand side (RHS) of the ODE~\eqref{e:O1system1} with respect to $\theta$;
(ii) zero-mean condition for the RHS of the ODE~\eqref{e:O1system2} with respect to $\theta$; and 
(iii) Fredholm solvability condition for the system~\eqref{e:O1system}.
These three conditions, which ensure that the corrections $u_1$ and $v_1$ to the leading order problem are periodic rather than growing in $\theta$,
yield the remaining three Whitham modulation equations.

Integrating the RHS of Eq.~\eqref{e:O1system2} and imposing that the mean over one period is zero yields Eq.~\eqref{eq:S3},
\be
\partial_xp = D_y\overline{u_0} - \overline{u_0}\partial_xq.   
\label{e:S3}   
\ee
Similarly, imposing the same condition on the RHS of Eq.~\eqref{e:O1system1} yields 
\be
\partial_t(\overline{u_0}) + \overline{F_0^{(0)}(u_0)} + \alpha\,\partial_y(q\overline{u_0}+p) = 0\,.
\label{e:Whitham4}
\ee
Equations~\eqref{e:S3} and~\eqref{e:Whitham4}
are two further Whitham modulation equations which are also common to all PDEs of the KP type
in the form of Eq.~\eqref{eq:1.1}.
The final Whitham modulation equation arises from the Fredholm solvability condition, which requires that the forcing term
in Eqs.~\eqref{e:O1system} be
orthogonal to the solutions of the adjoint of the homogeneous problem.
To write this condition,
one must threfore first study the homogeneous part of the system~\eqref{e:O1system}.
To this end, it is convenient to eliminate $v_1'$ in Eq.~\eqref{e:O1system1} using Eq.~\eqref{e:O1system2}, obtaining
\be
- \omega u_1' + F_1^{(0)}(u_0,u_1) + \alpha q^2 k u_1' = -G\,,
\label{e:ODEu1}
\ee
where
\be
G = \prt_tu_0 + F_0^{(0)}(u_0) + \alpha\,[\prt_y(qu_0+p) + q\prt_yu_0 - q\prt_x(qu_0 + p)]\,.
\ee
The Fredholm solvability condition is then simply
\be
\overline{w_\mathrm{adj}G} = 0\,,
\label{e:Fredholm}
\ee
where $w_\mathrm{adj}$ is the solution of the adjoint problem to~\eqref{e:ODEu1} with zero RHS.
To find~$w_\mathrm{adj}$, it is convenient to write the ODEs~\eqref{e:ODEu0} and~\eqref{e:ODEu1} respectively as
\be
L_0 u_0 = 0\,,\qquad L_1 u_1 = - G\,,
\ee
where 
\be
L_0w = F^{(-1)}(w) - \Omega \partial_\theta w\,,\qquad
L_1w = F_1^{(0)}(u_0,w) - \Omega \partial_\theta w\,.
\ee
Importantly, in all three specific examples considered, the multiple scales expansion imparts a certain structure on these two
differential operators, namely:
\be
F^{(-1)}(\,\cdot\,) = L\,\partial_\theta\,,\qquad
F_1^{(0)}(u_0,\,\cdot\,) = \partial_\theta L\,.
\label{e:L0vsL1}
\ee
Specifically:
\be
L_\mathrm{KdV} = k^3\partial_\theta^2 + 6ku_0\,,\qquad
L_\mathrm{BO} = k^2\partial_\theta H[\cdot] + ku_0\,,\qquad
L_\mathrm{mKdV} = k^3\partial_\theta^2 - 6ku_0^2\,.
\ee
Thus, in each case we can write the homogeneous problem at $O(1/\epsilon)$ and $O(1)$ respectively as
\be
(L - \Omega)\,\partial_\theta w = 0\,,\qquad 
\partial_\theta(L - \Omega) \,w = 0\,,\qquad 
\ee
Because of this structure, and the fact that $L$ is self-adjoint in each case, we have that the adjoint of $L_1$ is simply $L_1^\dag = - L_0$.
Thus, since the two periodic solutions of equation $L_0w = 0$ are a constant and $u_0$, and a constant yields already known condition $\overline G=0$, we take $w_\mathrm{adj} = u_0$, and the Fredholm solvability condition~\eqref{e:Fredholm} becomes 
\be
\overline{u_0\partial_t(u_0)} + \overline{u_0F^{(0)}_0(u_0)}
  + \overline{u_0(D_y(qu_0 + p) + q\partial_yu_0)} = 0\,.
\label{e:Whitham5}
\ee
Equation~\eqref{e:Whitham5} provides the last Whitham modulation equation.

\section{Conclusions}
\label{s:conclusions}

We have derived Whitham modulation equations for the KP, 2DBO and m2KP equations. From these modulation equations we derived hydrodynamic systems which are the analog of the (1+1)-dimensional Gurevich-Pitaevskii system. 
We also demonstrated how systems of Whitham modulation equations for (2+1)-dimensional PDEs of KP type can be derived in a unified way. 
For that part of the system related to the underlying Riemann variables, the derivation is in many respects similar to the derivation of its (1+1)-dimensional counterpart. 
The (2+1)-dimensional Whitham equations obtained here are richer than their one-dimensional reductions, and they can be expected to provide interesting new behavior and a larger variety of solutions.
We expect that the modulation systems obtained in this way will be useful in studying various physical phenomena such as DSWs and stability of periodic solutions. 

We point out that the form of the Fredholm solvability condition given in Eq.~\eqref{e:Whitham5} depends on the relation~\eqref{e:L0vsL1} between the homogeneous operators $L_0$ and $L_1$ 
which appear respectively at $O(1/\epsilon)$ and at $O(1)$ in the multiple scales expansion. 
Although this relation holds for the three specific PDEs considered in this work we expect this approach to be useful for many other systems.


It is also important to note that all the Whitham modulation systems that we have derived here comprise a closed system of equations for the variables $r_1,r_2,r_3,q$ and $p$.  
As such, these systems are interesting objects of study on their own right, without any need to go back to the original PDEs~\eqref{eq:1.1}.
Nonetheless, if one wants to use these systems to study the behavior of solutions of the original PDEs in the small dispersion limit, including the formation of DSWs,
the ICs for the dependent variables $r_1,r_2,r_3,q$ and $p$ should be chosen so that the constraint~\eqref{eq:kl} is satisfied at time zero.
For the Whitham systems for the KP and 2DBO equations, a prescription for doing so in order to consider (2+1)-generalizations of the 
Riemann problem for the corresponding one-dimensional systems was given in \cite{ABW-KP,ABW-BO}.
A similar procedure applies to the Whitham system for the m2KP equation.

\bigskip
{\bf\large Acknowledgments.} We are pleased to acknowledge help from Justin Cole with numerical simulations. We also thank Qiao Wang for many interesting conversations.
This work was partially supported by the National Science Foundation 
under grant numbers
DMS-1712793, 
DMS-1615524
and 
DMS-1614623.

\def\thesection{Appendix~\Alph{section}:}


\appendix
\section{Auxiliary formulas}
\label{a:auxiliary}

The first and second complete elliptic integrals $K=K(m)$ and $E=E(m)$ satisfy the following differential equations in $m$:
\be
K'(m) = \frac{E-(1-m)K}{2m(1-m)},   \qquad  E'(m) = -\frac{K-E}{2m},   \label{eq:K'E'}
\ee
from which it also follows e.g.~that
\be
\left(\frac{E}{K}\right)'(m) = \frac{1}{m}\left(\frac{E}{K} - \frac{1}{2} - \frac{E^2}{2(1-m)K^2}\right).   \label{eq:EK}
\ee
The third complete elliptic integral, cf.~e.g.~\cite{BF71}
\be
\Pi(\gamma,m) = \int_0^{K(m)}\frac{dz}{1-\gamma\;\text{sn}^2(z;m)},   \label{eq:Pi}
\ee
has the following derivatives with respect to its two arguments:
\bse
\begin{gather}
\frac{\prt\Pi(\gamma,m)}{\prt m} = \frac{E - (1-m)\Pi}{2(1-m)(m-\gamma)},   \label{eq:mPi}
\\
\frac{\prt\Pi(\gamma,m)}{\prt \gamma} = \frac{(m-\gamma^2)\Pi - \gamma E - (m-\gamma)K}{2\gamma(1-\gamma)(m-\gamma)}.   \label{eq:gPi}
\end{gather}
\ese





\par The KdV equation is known to have the following diagonal leading order Whitham equations~\cite{Whitham74} in Riemann variables $\{r_j\}$, $j=1,2, 3$,
\be
\prt_tr_j + v_j\prt_xr_j = 0,   \qquad   r_1<r_2<r_3,
\ee
or, equivalently, 
\be
Dr_j + \Delta_j\prt_xr_j = 0,   \qquad  v_j = V + \Delta_j = 2(r_1+r_2+r_3) + \Delta_j,   \label{eq:rW0}
\ee
where quantities $\Delta_j$ are given by general formula~\cite{GurKrEl92}
\be
\Delta_j = \frac{k}{\prt_jk}\prt_jV = \frac{2k}{\prt_jk},   \qquad  \prt_j \equiv \frac{\prt}{\prt r_j}.   \label{eq:Del-v}
\ee
Thus, they are determined by log-derivatives of $k$ with respect to the Riemann $r$-variables. These relations are obtained from definitions Eqs.~(\ref{eq:e123}), (\ref{eq:k2}), the first formula of Eq.~(\ref{eq:K'E'}) and Eq.~(\ref{eq:la-r}); they are given by Eq.~(\ref{eq:djk}). 

\par The Hilbert transform of a function $f(x)$ is defined as
\be
\mathcal{H}[f(x)] = \frac{1}{\pi} ~ \dashint_{-\infty}^\infty \frac{f(y)}{y-x} dy\,
\ee
where $\dashint$  denotes the Cauchy principal value integral. 



\section{``KdV-diagonalization" of the KP-Whitham system}

First, we form the combinations of Whitham equations $(\ref{eq:Q})-2Q\cdot(\ref{eq:kKP})$ and $(\ref{eq:P})-2P\cdot(\ref{eq:kKP})$ to get, respectively
\bse
\begin{gather}
\left(DQ - Q\frac{Dk}{k}\right) - \prt_xe_2 + \alpha(Y_1-QY_0) = 0,  \label{eq:tQ}
\\
\left(DP - P\frac{Dk}{k}\right) - 6\prt_xe_3 + \alpha(Y_2-PY_0) = 0.   \label{eq:tP}
\end{gather}
\ese
From Eq.~(\ref{eq:la-r}), we express the elementary symmetric functions of the roots $\lambda_j$ in terms of the power sums Eq.~(\ref{eq:pn}) of the $r$-variables,
\be
e_1= p_1,   \qquad  e_2 = p_1^2 - 2p_2,   \qquad  e_3 = 2p_1p_2 - \frac{p_1^3}{3} - \frac{8p_3}{3}.    \label{eq:e-p}
\ee
Now taking the combination $1/4\cdot(\ref{eq:tQ}) + p_1/2\cdot(\ref{eq:kKP})$ gives
\be
\frac{1}{4}\left(DQ - Q\frac{Dk}{k} + p_1\frac{Dk}{k}\right) + \frac{\prt_xp_2}{2} + \frac{\alpha}{4}(Y_1 + (p_1-Q)Y_0) = 0,     \label{eq:hQ}
\ee
and taking the combination $1/48\cdot(\ref{eq:tP}) + p_1/2\cdot(\ref{eq:hQ}) + (2p_2-p_1^2)/8\cdot(\ref{eq:kKP})$, gives
\begin{multline}
\frac{1}{48}\left(DP - P\frac{Dk}{k}\right) + \frac{p_1}{8}\left(DQ - Q\frac{Dk}{k}\right) + \frac{(2p_2+p_1^2)}{16}\frac{Dk}{k} + 
{ }\\{ }
+ \frac{\prt_xp_3}{3} + \frac{\alpha}{8}\left[ \frac{Y_2}{6} + p_1Y_1 + \left(\frac{2p_2+p_1^2-2p_1Q}{2} - \frac{P}{6}\right)Y_0 \right] = 0.   \label{eq:hP} 
\end{multline}
In terms of $r$-variables, 
\be
P = VQ+C_1 = 2p_1Q-p_1^2+2p_2,   \qquad  Q = r_1+r_2-r_3 + 2(r_3-r_1)\frac{E}{K},   \label{eq:QP}
\ee
and 
\be
\prt_jQ = (Q-p_1+2r_j)\frac{\prt_jk}{k},   \label{eq:dQ}
\ee
which is a consequence of Eqs.~(\ref{eq:QP}), (\ref{eq:djk}) and Eq.~(\ref{eq:EK}) of Appendix A. Upon using Eqs.~(\ref{eq:QP}), (\ref{eq:dQ}) and expressing everything in terms of $r$-variables and log-derivatives $\prt_jk/k$, Eqs.~(\ref{eq:kKP}), (\ref{eq:hQ}), (\ref{eq:hP}) are brought to the form, respectively,
\bse
\begin{gather}
\sum_{j=1}^3\left(\frac{1}{2}\frac{\prt_jk}{k}Dr_j + \prt_xr_j\right) + \alpha W_1= 0,    \label{eq:0W}
\\
\sum_{j=1}^3r_j\left(\frac{1}{2}\frac{\prt_jk}{k}Dr_j + \prt_xr_j\right) + \alpha W_2 = 0,    \label{eq:1W}
\\
\sum_{j=1}^3r_j^2\left(\frac{1}{2}\frac{\prt_jk}{k}Dr_j + \prt_xr_j\right) + \alpha W_3 = 0,    \label{eq:2W}
\end{gather}
\ese
where $W_1$, $W_2$ and $W_3$ are given by Eqs.~(\ref{eq:W1}), (\ref{eq:W2}), (\ref{eq:W3}), respectively. The system of Eqs.~(\ref{eq:0W}), (\ref{eq:1W}), (\ref{eq:2W}) can be rewritten in matrix-vector form as
\be
\Delta_{jl}\left(\frac{1}{2}\frac{\prt_lk}{k}Dr_l + \prt_xr_l\right) + \alpha W_j = 0,  \qquad  j,l = 1,2,3,   \label{eq:MVW}
\ee
where 
$$
\Delta = \left( \begin{array}{ccc} 1 & 1 & 1 \\ r_1 & r_2 & r_3 \\ r_1^2 & r_2^2 & r_3^2 \end{array}  \right)  
$$
is the Vandermonde matrix. Multiplying Eq.~(\ref{eq:MVW}) on the left by the inverse of the Vandermonde matrix, we obtain the three equations Eq.~(\ref{eq:rjg}) diagonal in the derivatives $\prt_tr_j$ and $\prt_xr_j$. Finally, using Eqs.~(\ref{eq:W1}), (\ref{eq:W2}), (\ref{eq:W3}) and expressions 
\be
Q = r_l + r_m - r_j + 4(r_j-r_l)(r_j-r_m)\frac{\prt_jk}{k} = p_1 - 2r_j + 4\left(3r_j^2 - 2p_1r_j + \frac{p_1^2-p_2}{2}\right)\frac{\prt_jk}{k},   \label{eq:Qr}
\ee
in Eq.~(\ref{eq:gj}), we bring the Whitham Eqs.~(\ref{eq:rjg}) to their final explicit form, see Eq.~(\ref{eq:rj}) of section 1.

\section{``mKdV-diagonalization" of the m2KP-Whitham system}

First, we make the combinations of Whitham equations $(\ref{eq:mQ})-Q\cdot(\ref{eq:mk})$ and $(\ref{eq:mQ2})-Q_2\cdot(\ref{eq:mk})$ to get, respectively
\bse
\begin{gather}
\left(DQ - Q\frac{Dk}{k}\right) + \prt_xA_1 + \alpha(Y_1-QY_0) = 0,  \label{eq:tQm}
\\
\left(DQ_2 - Q_2\frac{Dk}{k}\right) - \prt_xA_2 + \alpha(Y_2-Q_2Y_0) = 0.   \label{eq:tQ2}
\end{gather}
\ese
Using Eq.~(\ref{eq:e-R}), we observe that derivatives $\prt_xr_j$, $j=1,2,3$, in Eqs.~(\ref{eq:mk}), (\ref{eq:tQm}), (\ref{eq:tQ2}) have coefficients polynomial in $R$-variables. So we diagonalize these equations with respect to these $x$-derivative terms. Let $\{i,l,m\}=\{1,2,3\}$ be the three different subscripts for $r_j$, $j=1,2,3$. Now taking the combination of equations $r_i\cdot(\ref{eq:tQm}) + r_lr_m\cdot(\ref{eq:mk})$ gives
\begin{multline}
r_i\left(DQ - Q\frac{Dk}{k}\right) + r_lr_m\frac{Dk}{k} + 4[(r_i^2-r_l^2)r_m\prt_xr_l + (r_i^2-r_m^2)r_l\prt_xr_m] + 
\\
+ \alpha(r_i(Y_1-QY_0) + r_lr_mY_0) = 0,     \label{eq:Qlm}
\end{multline}
and taking the combination of equations $(\ref{eq:tQ2}) + (r_l^2+r_m^2-r_i^2)\cdot(\ref{eq:mk})$ gives
\begin{multline}
\left(DQ_2 - Q_2\frac{Dk}{k}\right) + (r_l^2+r_m^2-r_i^2)\frac{Dk}{k} + 8[(r_i^2-r_l^2)r_l\prt_xr_l + (r_i^2-r_m^2)r_m\prt_xr_m] +
{ }\\{ }
+ \alpha\left[ Y_2 - Q_2Y_0 + (r_l^2+r_m^2-r_i^2)Y_0 \right] = 0.   \label{eq:Q2lm}
\end{multline}
In Eqs.~(\ref{eq:Qlm}), (\ref{eq:Q2lm}) we have only two of the three $\prt_xr_j$-derivative terms left. Now we combine them in $r_m\cdot(Q2lm) - 2r_l\cdot(Qlm)$ to get
\begin{multline}
r_m\left(DQ_2 + (r_m^2-r_l^2-r_i^2 - Q_2)\frac{Dk}{k}\right) - 2r_lr_i\left(DQ - Q\frac{Dk}{k}\right) - 8(r_m^2-r_l^2)(r_m^2-r_i^2)\prt_xr_m +
{ }\\{ }
+ \alpha\left[ r_mY_2 - 2r_lr_iY_1 + (2r_lr_iQ - r_mQ_2 + r_m(r_m^2-r_l^2-r_i^2))Y_0 \right] = 0,   \label{eq:Rm}
\end{multline}
equation containing only $\prt_xr_m$-derivative term. It must be diagonal also in the $D$-derivative terms coming from mKdV. To see this, we use the identities
\be
\prt_mQ - Q\frac{\prt_mk}{k} = -\frac{r_lr_i}{r_m}\frac{\prt_mk}{k},   \label{eq:IdQ}
\ee
\be
\prt_mQ_2 - Q_2\frac{\prt_mk}{k} = (r_m^2-r_l^2-r_i^2)\frac{\prt_mk}{k},   \label{eq:IdQ2}
\ee
which can be derived from the explicit expressions Eqs.~(\ref{eq:mgR}), (\ref{eq:QR}) in terms of elliptic functions and the equations for their derivatives Eqs.~(\ref{eq:mPi}), (\ref{eq:gPi}) of Appendix A. Then we obtain another two identities,
\be
r_m\left(\prt_iQ_2 - Q_2\frac{\prt_ik}{k} + (r_m^2-r_l^2-r_i^2)\frac{\prt_ik}{k}\right) - 2r_lr_i\left(\prt_iQ - Q\frac{\prt_ik}{k}\right) = 0,   \quad  i\neq l\neq m\neq i,   \label{eq:Ida}
\ee
which implies vanishing of the non-diagonal $D$-derivative terms in Eq.~(\ref{eq:Rm}), and 
\begin{multline}
r_m\left(\prt_mQ_2 - Q_2\frac{\prt_mk}{k} + (r_m^2-r_l^2-r_i^2)\frac{\prt_mk}{k}\right) - 2r_lr_i\left(\prt_mQ - Q\frac{\prt_mk}{k}\right) = 
{ }\\{ }
= \frac{2(r_m^2-r_i^2)(r_m^2-r_l^2)}{r_m}\frac{\prt_mk}{k},   \label{eq:Idd}
\end{multline}
which allows us to rewrite Eq.~(\ref{eq:Rm}) as
\be
Dr_m - \frac{4r_m\prt_xr_m}{\prt_mk/k} + \alpha\frac{r_m\left[ r_mY_2 - 2r_lr_iY_1 + (2r_lr_iQ - r_mQ_2 + r_m(r_m^2-r_l^2-r_i^2))Y_0 \right]}{2(r_m^2-r_i^2)(r_m^2-r_l^2)\prt_mk/k}.   \label{eq:Rm-d}
\ee
Now we recall Eq.~(\ref{eq:YsMK}) and express
\bse
\begin{gather}
Y_0 = q^2\sum_j\frac{\prt_jk}{k}\prt_xr_j + 2q\sum_j\frac{\prt_jk}{k}D_yr_j,  \label{eq:mY0}
\\
Y_1 = q^2\sum_j\prt_jQ\prt_xr_j + 2q\sum_j\prt_jQD_yr_j + QD_yq+D_yp,   \label{eq:mY1}
\\
Y_2 = q^2\sum_j\prt_jQ_2\prt_xr_j + 2q\sum_j\prt_jQ_2D_yr_j + 2Q_2D_yq + 2QD_yp.   \label{eq:mY2}
\end{gather}
\ese
To simplify the last term in Eq.~(\ref{eq:Rm-d}), we use Eqs.~(\ref{eq:mY0}), (\ref{eq:mY1}), (\ref{eq:mY2}) and the identities
\be
\prt_jQ = \left(Q - \frac{r_1r_2r_3}{r_j^2}\right)\frac{\prt_jk}{k},   \qquad  \prt_jQ_2 = (Q_2 + 2r_j^2 - r_1^2 - r_2^2 - r_3^2)\frac{\prt_jk}{k}.    \label{eq:mdQ}
\ee
which are consequences of Eqs.~(\ref{eq:kjR}), (\ref{eq:QR}) and Eqs.~(\ref{eq:EK}), (\ref{eq:mPi}) and (\ref{eq:gPi}) of Appendix A. This way we finally obtain Eqs.~(\ref{eq:Rj}) of the main text.

\section{Gurevich-Pitaevskii problem for the defocusing mKdV equation}
\label{s:mKdV}

When $\alpha=0$, the Whitham equations~\eqref{e:m2KP-Whitham} for the m2KP equation reduce to those for the mKdV equation, namely to the three equations~\eqref{eq:Rj}, which then read
\be
\prt_tr_j + v_j \prt_xr_j = 0.   \label{eq:WmKdV}
\ee
A natural step in preparation for the study of the m2KP equation is to look for (and find) an analog of the seminal Gurevich-Pitaevskii KdV solution~\cite{GurPit73} for the mKdV equation. We note that DSWs for the mKdV equation were first studied in \cite{Bik93}. Recently, a general classification of DSWs and rarefaction waves arising from an initial step for the mKdV equation was presented in~\cite{ElHSh17,KamEtAl12},
while \cite{March08} studied the evolution of initial step for the focusing mKdV equation.

Here, however, we give full analytical details of the DSW solution for the defocusing mKdV equation in a simple case, which is the counterpart of the Riemann problem for the KdV equation studied in~\cite{GurPit73}.
To do this, we consider the following piecewise constant initial condition (IC) for $u(x,t)$:
\be
u(x,0) = \begin{cases} 0, & x<0,\\ 1, & x>0. \end{cases}
\label{eq:IC}
\ee
It is relevant to consider such a step-up IC since we have seen that $V<0$ for all real-valued solutions of the m2KP equation [cf.\ Eq.~\eqref{eq:e-R}]. 
This IC implies the corresponding ICs for the Riemann invariants $r_j$,
\be
r_1(x,0) = 0,  \qquad r_3(x,0) = 1,   \qquad r_2(x,0) =  \left\{ \begin{array}{cc} 1, & x<0 \\ 0, & x>0. \end{array}  \right.   \label{eq:ICR}
\ee
The above step IC for the mKdV equation is not the most general, and the corresponding solution, described below, is the borderline case between the single cnoidal DSW solutions realized when $0{}<{} r_1(x,0)<r_3(x,0)$ and more complicated solutions realized when the values of $u(x,0)$ on two sides of the step are of the opposite sign, i.e., $r_1(x,0)<0<r_3(x,0)$~\cite{ElHSh17, KamEtAl12}.

The cnoidal-wave solution~\eqref{eq:u0R}
satisfies  Eq.~(\ref{eq:IC}) e.g.~for $\theta_*=0$. 
Indeed, at $t=0$, for $x<0$, we have $r_2=r_3$, $m=0$ and $u_0 = r_1=0$. On the other hand, for $x>0$, we have $r_2=r_1$, so $m=1$ and $u_0=r_3=1$. 
Looking for a self-similar solution of the Whitham equations, namely, $r_j=r_j(\xi)$, $\xi=x/t$, we find, as in the case of KdV, that
\be
r_1(\xi) \equiv 0,  \qquad r_3(\xi) \equiv 1,  \qquad  v_2 = v_2(r_2) = \xi.   \label{eq:mGP}
\ee
Let $r_2 = s = s(\xi)$. Then $m=1-s^2$, $\gamma=1-s$ and the solution for $s(\xi)$ in the oscillation domain where $0<s<1$ is implicitly given by formula  
\be
\xi = v_2(s) = -2(1+s^2) - \frac{4s^2(1-s^2)}{E(m)/K(m) - s^2}.   \label{eq:sGP}
\ee
In the domain $0<s<1$, the solution $u_0$ of Eq.~(\ref{eq:u0R}) oscillates with maxima $u_{max} = r_1+r_3-r_2=1-s$ and minima $u_{min}=r_1-(r_3-r_2)=-(1-s)$. The highest maximum is at the right (trailing) edge where $s=0$ and $u_0=1$ while the deepest minimum is just to the left of the trailing edge with $u$ approximately equal to $-1$. The function $v_2(s)$ determines the leading and trailing edge velocities of the DSW for the mKdV equation. At the leading (left) edge $s=1$ and we get $v_2=v_2(1)=-12$, and at the trailing (right) edge $s=0$ and $v_2=v_2(0)=-2$. Thus, the DSW structure is expanding to the left of the initial jump, with both leading and trailing edge velocities negative. A typical DSW is given in Fig.~\ref{f:m2KP}(left). 
This is similar to the DSW for the one-dimensional (1D) NLS equation, where both velocities have the same sign, and is different from the DSW for KdV, where their signs are opposite. The DSW described by the function $u_0(x,t)$ of Eq.~(\ref{eq:u0R}) is shown in Fig.~1 for $\epsilon=0.05$ at a typical time $t=0.5$. At its trailing (right) edge ``dark solitons" are seen forming; this is similar to the DSW that arises in the defocusing 1D NLS equation, see e.g.~\cite{GBYK06,ElHoefRev16,HoeferEtAl06,YK99} and references therein.



\bigskip
\begin{thebibliography}{10}
\advance\itemsep-5pt

\bibitem{ABW-KP}
M.J. ~Ablowitz, G.~Biondini, Q.~Wang.
\newblock Whitham modulation theory for the Kadomtsev-Petviashvili equation.
\newblock {\em Proc. Roy. Soc. A}, 473:20160695, 2017.

\bibitem{ABW-BO}
M.J.~Ablowitz, G.~Biondini, Q.~Wang.
\newblock Whitham modulation theory for the two-dimensional Benjamin-Ono equation.
\newblock {\em Phys. Rev. E} 96:032225, 2017.

\bibitem{ADM}
M.J.~Ablowitz, A.~Demirci, Y.-P.~Ma.
\newblock Dispersive shock waves in the Kadomtsev-Petviashvili and two dimensional Benjamin-Ono equations.
\newblock {\em Physica D}, 333:84--98, 2016.

\bibitem{AbSeg81}
M.J.~Ablowitz, H.~Segur.
\newblock Solitons and the inverse scattering transform.
\newblock {\em SIAM}, Philadelphia, 1981.

\bibitem{AbSeg80}
M.J. Ablowitz and H. Segur,
\newblock Long internal waves in fluids of great depth, 
\newblock {\it Stud. Appl. Math.}, 62:249--262, 1980.

\bibitem{Bik93}
R.~Bikbaev.
\newblock Shock waves in one-dimensional models with cubic nonlinearity.
\newblock {\em Theor. Math. Phys.}, 97:1236--1249, 1993.

\bibitem{GBYK06}
G.~Biondini and Y. Kodama, 
\newblock On the Whitham equations for the defocusing nonlinear Schr\"o\-dinger equation with step initial data,
\newblock {\em J. Nonlin.\ Sci.} 16:435--481, 2006.

\bibitem{Bog91}
V.~Bogaevskii.
\newblock On Korteweg-de Vries, Kadomtsev-Petviashvili and Boussinesq equations in the theory of modulations.
\newblock {\em Zh. Vychisl. Mat. i Mat. Fiz.}, 30:1487--1501, 1990.

\bibitem{BF71}
P.~Byrd, M.~Friedman.
\newblock Handbook of elliptic integrals for engineers and scientists.
\newblock {\em Springer-Verlag}, 2nd edition, Berlin, 1971.

\bibitem{DN-mKdV}
C.~Driscoll, T.~O'Neil.
\newblock Modulational instability of cnoidal wave solutions of the modified Korteweg-de Vries equation.
\newblock {\em J. Math. Phys.}, 17:1196--1200, 1976.

\bibitem{DuGrKl16}
B.~Dubrovin, T.~Grava, C.~Klein.
\newblock On critical behaviour in generalized Kadomtsev-Petviashvili equations.
\newblock {\em Physica D}, 133:157--170, 2016.

\bibitem{GurKrEl92}
A.~Gurevich, A.~Krylov, G.~El.
\newblock Evolution of a Riemann wave in dispersive hydrodynamics.
\newblock {\em J. Exp. Theor. Phys.}, 74:957--962, 1992.

\bibitem{GurPit73}
A.~Gurevich, L.~Pitaevskii.
\newblock Nonstationary structure of a collisionless shock wave.
\newblock {\em J. Exp. Theor. Phys.}, 38:291--297, 1974.


\bibitem{ElHoefRev16}
G.~El, M.~Hoefer.
\newblock Dispersive shock waves and modulation theory.
\newblock {\em Phys. D}, 333:11--65, 2016.

\bibitem{ElHSh17}
G.~El, M.~Hoefer, M.~Shearer.
\newblock Dispersive and diffusive-dispersive shock waves for nonconvex conservation laws.
\newblock {\em SIAM Review}, 59:3--61, 2017.

\bibitem{Grava17}
T.~Grava.
\newblock Whitham modulation equations and application to small dispersion asymptotics and long time asymptotics of nonlinear dispersive equations.
\newblock {\em arXiv:1701.00069}, 2017.

\bibitem{GrKlPi17}
T.~Grava, C.~Klein, G.~Pitton.
\newblock Numerical study of the Kadomtsev-Petviashvili equation and dispersive shock waves.
\newblock {\em arXiv:1706.04104}, 2017.

\bibitem{HoeferEtAl06}
M.~Hoefer, M.~Ablowitz, I.~Coddington, E.~Cornell, P.~Engels, and V.~Schweikhard.
\newblock Dispersive and classical shock waves in Bose-Einstein condensates and gas dynamics.
\newblock {\em Phys. Rev. A}, 74:023623, 2006.

\bibitem{InfRow}
E.~Infeld, G.~Rowlands.
\newblock Nonlinear waves, solitons and chaos.
\newblock {\em Cambridge Univ. Press}, Cambridge, UK, 2000.

\bibitem{JM83}
M.~Jimbo, T.~Miwa.
\newblock Solitons and infinite-dimensional Lie algebras.
\newblock {\em Publ. RIMS}, 19:943--1001, 1983.

\bibitem{KamEtAl12}
A.~Kamchatnov, Y.~Kuo, T.~Lin, T.~Horng, S.~Gou, R.~Clift, G.~El, and R.~Grimshaw.
\newblock Undular bore theory for the Gardner equation.
\newblock {\em Phys. Rev. E}, 86:036605, 2012.


\bibitem{YK99}
Y. Kodama, 
\newblock The {Whitham} equations for optical communications: Mathematical theory of {NRZ},
\newblock {\em SIAM J. Appl. Math.} 59:2162--2192, 1999.

\bibitem{Kon82}
B.~Konopelchenko.
\newblock On the gauge-invariant description of the evolution equations integrable by Gelfand-Dikij spectral problems.
\newblock {\em Phys. Lett. A}, 92:323--327, 1982.

\bibitem{Kr88}
I.~Krichever.
\newblock Method of averaging for two-dimensional ``integrable" equations.
\newblock {\em Funct. Anal. Appl.}, 22:37--52, 1988.

\bibitem{March08}
T.~Marchant.
\newblock Undular bores and the initial-boundary value problem for the modified Korteweg-de Vries equation.
\newblock {\em Wave Motion}, 45:540--555, 2008.

\bibitem{Mats98}
Y.~Matsuno.
\newblock Nonlinear modulation of periodic waves in the small dispersion limit of the Benjamin-Ono equation.
\newblock {\em Phys. Rev. E}, 58:7934--7940, 1998.

\bibitem{TuFa85}
S.~Turitsyn, G.~Falkovich.
\newblock Stability of magneto-elastic solitons and self-focusing of sound in antiferromagnets.
\newblock {\em J. Exp. Theor. Phys.}, 62:146--152, 1985.

\bibitem{WAS94}
X.~P.~Wang, M.J.~Ablowitz, H.~Segur.
\newblock Wave collapse and instability of solitary waves of a generalized Kadomtsev-Petviashvili equation.
\newblock {\em Physica D}, 333:84--98, 1994.

\bibitem{Whitham65}
G.~Whitham.
\newblock Nonlinear dispersive waves.
\newblock {\em Proc. Roy. Soc.}, 283:238--261, 1965.

\bibitem{Whitham74}
G.~Whitham.
\newblock Linear and nonlinear waves.
\newblock {\em Wiley}, NY, 1974.

\end{thebibliography}
\end{document}